\newif\ifconfver
\confverfalse      

\ifconfver
     \documentclass[10pt,twocolumn,twoside]{IEEEtran}
\else
    \documentclass[11pt,draftcls,onecolumn]{IEEEtran}
\fi

\usepackage{calc,amsfonts,amssymb,amsmath,bm,url,color,theorem,cite}
\usepackage{graphicx,subfigure}
\usepackage{epsfig,multirow}
\usepackage{algorithmic,algorithm}

\newlength{\twidth}
\ifconfver
\else
\fi

    \makeatletter
    \def\multilimits@{\bgroup
  \Let@
  \restore@math@cr
  \default@tag
 \baselineskip\fontdimen10 \scriptfont\tw@
 \advance\baselineskip\fontdimen12 \scriptfont\tw@
 \lineskip\thr@@\fontdimen8 \scriptfont\thr@@
 \lineskiplimit\lineskip
 \vbox\bgroup\ialign\bgroup\hfil$\m@th\scriptstyle{##}$\hfil\crcr}
    \def\Sb{_\multilimits@}
    \def\endSb{\crcr\egroup\egroup\egroup}
\makeatother

\newtheorem{Lemma}{Lemma}
\newtheorem{Prop}{Proposition}
\newtheorem{Theorem}{Theorem}

\newtheorem{Fact}{Fact}
\theorembodyfont{\rmfamily}

\definecolor{orange}{RGB}{255,107,0}
\definecolor{purple}{rgb}{0.627,0.125,0.941}

{\begin{list}{}%
    {\setlength{\rightmargin}{0pc}%
    \setlength{\leftmargin}{1pc} }} %
{\end{list}}

{\begin{list}{}%
    {\setlength{\rightmargin}{1pc}%
    \setlength{\labelsep}{0pc}
    \setlength{\leftmargin}{1pc} }} %
{\end{list}}

{\begin{list}{}%
    {\setlength{\rightmargin}{0pc}%
    \setlength{\leftmargin}{0pc} }} %
{\end{list}}

{\begin{list}{}{
    \settowidth{\labelwidth}{\mbox{\textnormal{#1}}}%
    \setlength{\leftmargin}{\labelwidth+\labelsep}}}%
{\end{list}}

\begin{document}

\bibliographystyle{IEEEtran}

\title{Physical-Layer Multicasting by Stochastic Transmit Beamforming and Alamouti Space-Time Coding}

\ifconfver \else {\linespread{1.1} \rm \fi

\author{
Sissi Xiaoxiao Wu$^\dag$, Wing-Kin Ma$^\ddag$, and Anthony Man-Cho So$^\S$
\ifconfver

\else
    \thanks{Copyright (c) 2012 IEEE. Personal use of this material is permitted. However, permission to use this material for any other purposes must be obtained from the IEEE by sending a request to pubs-permissions@ieee.org.}
\fi
\thanks{This work was supported in part by the Hong Kong Research Grant Council (RGC) General Research Fund (GRF) Project CUHK 416012, and in part by The Chinese University of Hong Kong Direct Grant No. 2050506.
 Part of this work was presented at ICASSP 2011 and ICASSP 2012.}
\thanks{$^\dag$Sissi Xiaoxiao Wu is with the
Department of Electronic Engineering, The Chinese University of Hong
Kong, Shatin, Hong Kong S.A.R., China. E-mail: xxwu@ee.cuhk.edu.hk.}
\thanks{$^\ddag$Wing-Kin Ma is the corresponding author. Address:
Department of Electronic Engineering, The Chinese University of Hong
Kong, Shatin, Hong Kong S.A.R., China. E-mail: wkma@ieee.org.}
\thanks{$^\S$Anthony Man-Cho So is with the
Department of Systems Engineering and Engineering Management, and, by courtesy, Department of Computer Science and Engineering and the CUHK-BGI Innovation Institute of Trans-omics, The Chinese University of Hong
Kong, Shatin, Hong Kong S.A.R., China. E-mail: manchoso@se.cuhk.edu.hk.}
}

\maketitle

\ifconfver \else
\begin{center} \vspace*{-3\baselineskip}
Final Version,
May 2013
\\[\baselineskip]
\end{center}
\fi

\begin{abstract}
Consider transceiver designs in a multiuser multi-input single-output (MISO) downlink channel, where the users are to receive the same data stream simultaneously.
This problem, known as physical-layer multicasting, has drawn much interest.
Presently, a popularized approach is transmit beamforming,
in which the beamforming optimization is handled by a rank-one approximation method called semidefinite relaxation (SDR).
SDR-based beamforming has been shown to be promising for a small or moderate number of users.
This paper describes two new
transceiver strategies for physical-layer multicasting.
The first strategy, called stochastic beamforming (SBF), randomizes the beamformer in a per-symbol time-varying manner,
so that the rank-one approximation in SDR can be bypassed.
We propose several efficiently realizable SBF schemes,
and prove that their multicast achievable rate gaps with respect to the MISO multicast capacity must be no worse than $0.8314$~bits/s/Hz,
irrespective of any other factors such as the number of users.
The use of channel coding and the assumption of sufficiently long code lengths play a crucial role in achieving the above result.
The second strategy combines transmit beamforming and the Alamouti space-time code.
The result is a rank-two generalization of SDR-based beamforming.
We show by analysis that this SDR-based beamformed Alamouti scheme has a better worst-case effective signal-to-noise ratio (SNR) scaling, and hence a better multicast rate scaling, than SDR-based beamforming.
We further the work by combining SBF and the beamformed Alamouti scheme,
wherein an improved constant rate gap of $0.39$~bits/s/Hz is proven.
Simulation results show that under a channel-coded, many-user setting,
the proposed multicast transceiver schemes yield significant SNR gains over SDR-based beamforming at the same bit error rate level.
%
\\\\
\noindent {\bfseries Index terms}$-$ physical-layer multicasting, multicast capacity,
transmit beamforming, semidefinite relaxation, semidefinite programming
\\[.5\baselineskip]
\ifconfver  \else
\noindent
{\bfseries EDICS}: MSP-CODR (MIMO precoder/decoder design),
MSP-STCD  (MIMO space-time coding and capacity),
MSP-CAPC  (MIMO capacity and performance)
\fi
\end{abstract}

%

\ifconfver \else } \fi

\ifconfver
\else
\newpage
\fi



\section{Introduction}
\label{sec:intro}

In recent years, the explosive growth in the demand for various wireless data services
has motivated a vast amount of research on
resource-efficient techniques for massive content delivery.
One scenario that has received significant attention is physical-layer multicasting, in which a
base station broadcasts
common information
to
a prespecified group of
users.
For instance, in the long term evolution (LTE) standard, a particular work item called multimedia broadcast multicast service (MBMS)~\cite{LTEbook} is being actively considered as a preferred mass media streaming option.

The scenario of interest in this paper is that of physical-layer multicasting in multiuser multiple-input single-output (MISO) downlink, assuming channel state information at the transmitter side (CSIT).
A central problem in this context is to develop efficient and physically realizable transceiver techniques.
Currently, a popularized approach is multicast beamforming, in which the physical-layer transmit strategy is single-stream beamforming, and the beamformer is designed so that users can simultaneously receive good quality of service (QoS).
The idea of using beamforming as a transmit strategy for physical-layer multicasting can be traced back to a 1998 paper by Narula {\em et al.}~\cite{Jnl:Narula_Multicast_BF98},
although more appropriate beamforming optimization formulations, namely,
the QoS-constrained problem and the max-min-fair problem, appeared later~\cite{Conferece:BF_Sun_Liu_04,Conferece:Sidiropoulos04}.
As it turns out, both of these formulations are NP-hard in general~\cite{MulticastSidiropoulos06,MulticastLuo07}.  To circumvent such intractability, a state of the art approach is semidefinite relaxation (SDR)~\cite{MulticastSidiropoulos06}.
The main observation behind this approach is that the beamforming problem can be reformulated as a rank-one constrained semidefinite program (SDP).  Thus, by dropping the non-convex rank constraint, one obtains a convex and tractable SDR problem, whose solution can then be used to generate a rank-one approximate solution to the original beamforming problem.
The viability of SDR-based multicast beamforming has been proven by both empirical evidence~\cite{MulticastSidiropoulos06} and theoretical analysis~\cite{MulticastLuo07,Conferece:CapacityLimitsLuo06}.
In fact, SDR-based multicast beamforming has sparked much interest in the area,
where we have seen the same fundamental idea of SDR being applied to many different beamforming scenarios;
see, e.g., the references in \cite{inbook:Tsung_Hui_bokcha}.
In addition,
we should note that the significance of multicast beamforming as demonstrated through SDR has motivated the development of many other competing optimization methods~\cite{Jnl:channel_orth_10,Conferece:ICC_Hunger_sucessive_BF_07,Jnl:successive_BF_2011,Jnl:SOCP_Bornhorst_12,Jnl:AO_Zhu_Prasad_12}.

While our main interest is in multicast transceiver designs under CSIT,
we should also briefly mention the no-CSIT case.
A common transmit strategy without CSIT is to transmit isotropically,
which is called the open-loop strategy
in the literature
and may physically be implemented by space-time coding~\cite{Conferece:BF_Sun_Liu_04}.
Open-loop system capacity analyses have been considered in \cite{Conferece:CapacityLimitsLuo06,Jnl:Subsection_Love_08}.
The work~\cite{Jnl:Subsection_Love_08} also considers antenna subset selection for striking a balance between the full CSIT and no CSIT settings.
Moreover, in \cite{Conferece:Wang_Love_06}, a diagonally precoded extension of the space-time coding approach was proposed for the full CSIT case.


\subsection{Motivations and Contributions}

The now popularized SDR-based multicast beamforming scheme has been shown to be capable of providing accurate approximations in a variety of practical regimes, most notably, in the cases where there is a small or moderate number of users.
However, the analyses in~\cite{MulticastLuo07} also reveal an inherent limitation, namely, the SDR approximation accuracy may degrade as the number of users increases.
The focus of this work is to pursue alternative physical-layer multicasting strategies that can deliver good performance even in the presence of large number of users.
Our endeavor is motivated from an information theoretic perspective---the SDR solution under the max-min-fair formulation is equivalent to the optimal transmit covariance of the multicast capacity~\cite{MulticastSidiropoulos06,Conferece:CapacityLimitsLuo06}.
Hence, instead of extracting a rank-one approximate solution from the SDR solution, which is the case in multicast beamforming,
we consider altering the transmit structure to embrace the non-rank-one nature of the multicast optimal transmit covariance (or the SDR solution).
Specifically, we propose two new physical-layer strategies for multicasting in this paper.

\subsubsection{Stochastic Beamforming}
The first strategy, called {\it stochastic beamforming (SBF)}, is to employ a per-symbol time variant, randomly generated, beamformer.  The underlying intuition of SBF is to use time-varying spatial randomizations to mimic the multicast capacity-optimal transmit covariance, thus producing ``rank-$r$ beamforming'' in a {\it virtual} manner for any $r\ge1$.
A distinguishing characteristic of our SBF framework is that channel coding (which is usually present in practical systems) is utilized to approach some kind of ergodic achievable rate metric.
We will
develop three efficiently and practically implementable schemes under the SBF framework.  Numerical simulations show that they can have significant bit-error-rate (BER) performance gains over SDR-based beamforming.  On the theoretical side, we prove that the achievable rate gaps of the proposed SBF schemes with respect to (w.r.t.) the multicast capacity must be no worse than $0.8314$~bits/s/Hz, {\it irrespective of any factors such as the number of users}.  From a practical viewpoint, this implies that even when there is a large number of users, SBF can still perform reasonably well.


\subsubsection{Alamouti-Assisted Rank-Two Beamforming}

Our second strategy is to develop
rank-$2$ generalizations of beamforming, both fixed and stochastic,
through the use of the Alamouti space-time code.
To motivate this strategy, we should first note that in the point-to-point multiple-input multiple-output (MIMO) literature, there has been interest in combining beamforming and space-time coding (STC) to provide rank-$r$ beamforming; see, e.g.,~\cite{Jongren_Ottersten02,Pascual_Palomar06,Mai_Paulraj06}.
However, developing a combined beamforming and STC scheme is a scenario-dependent challenge, as evidenced in the above referenced work.
The reason is that many available space-time codes are designed for performance metrics in point-to-point CSIT-uninformed scenarios, such as diversity order or diversity multiplexing tradeoff,
and those merits do not always carry forward to another MIMO scenario that has a different performance metric.
In Section~4.2 of the companion technical report~\cite{CompTechRep},
we provide simulation results that demonstrate a direct combination of beamforming and STC based on intuition would lead to poor performance in the multicast scenario.

There is however an exception where beamformed STC designs can be tractable, namely, when the class of orthogonal space-time block codes (OSTBCs) is used.
OSTBCs are well known to be simple to implement, and, more importantly, their performance can be easily characterized by an explicit signal-to-noise ratio (SNR) expression.
Some representative point-to-point beamformed STC designs are, in fact, based on OSTBCs~\cite{Jongren_Ottersten02,Pascual_Palomar06,Mai_Paulraj06}.
On the other hand, one must note that full-rate OSTBCs do not exist for dimensions higher than two~\cite{Liang_OSTBC_03}.


In view of the above discussion, we will consider beamformed STC based on the two-dimensional full-rate OSTBC, that is, the well-known Alamouti space-time code.
We first develop an SDR-based fixed beamformed Alamouti scheme, which is a rank-two generalization of the previous (rank-one) SDR-based beamforming framework.  Our analysis shows that in terms of the effective worst-user SNR, the worst-case approximation accuracy of the beamformed Alamouti scheme degrades only at a rate of $\sqrt{M}$, where $M$ is the number of users.  This is an improvement over the previous beamforming scheme, where the provable worst-case approximation accuracy degrades at the higher rate of $M$~\cite{MulticastLuo07}.  Next, we combine the SBF strategy and the beamformed Alamouti scheme; that is, we produce virtually rank-$r$ beamforming from physically rank-two beamforming.
By analysis, we show that the SBF Alamouti schemes have a worst-case multicast achievable rate gap of $0.39$~bits/s/Hz, which is better than the previous $0.8314$ bits/s/Hz bound for the SBF schemes.
The SBF Alamouti schemes also yield the best coded BER performance by simulations when compared to beamforming and other proposed schemes.

\subsection{Related Works}

We should
mention some existing works that might seem related to SBF, and contrast the differences.
At first sight, using randomness in beamforming may remind one of
the opportunistic beamforming (OBF) technique~\cite{Jnl:OpportunisticTse02}.
However, OBF deals with user scheduling in a multiuser TDMA setting, which is
a very different scenario from multicasting.
By closely examining OBF and SBF, one would find that the ways randomness is used also have much difference: OBF is a per-frame randomized approach without CSIT, while SBF is per-symbol random with CSIT.
For a similar reason, SBF is different from the randomized space-time coding approach for cooperative communication~\cite{Jnl:randomized_space_time_coding07}---the latter is per-frame random without CSIT, with an aim to harvest cooperative diversity.
Moreover, it is interesting to note that the philosophical possibility of randomizing the beamfomer was vaguely alluded to in a study of the unicast scenario \cite{Bengtsson2001}, although no further investigation was provided.
In fact, the authors there never needed to---they showed that in unicasting, SDR always has a rank-one solution, i.e., transmit beamforming is sufficient in unicasting.
However, this result does not apply to multicasting~\cite{MulticastSidiropoulos06,MulticastLuo07}.
In this study, the idea of utilizing channel coding, and the subsequent ergodic rate characterization for multicasting, are new.

We should also describe related work on our fixed beamformed Alamouti scheme.
As mentioned earlier,
the beamformed Alamouti structure, or, more generally, the beamformed OSTBC structure,
has previously appeared in the point-to-point MIMO literature, e.g., \cite{Jongren_Ottersten02,Pascual_Palomar06,Mai_Paulraj06}.
Also, in the multicast scenario, there is an early work  \cite{Conferece:Wang_Love_06}
where the authors considered a diagonally precoded OSTBC scheme with per-antenna power allocation (rather than beamforming).
The issue that is different in the present scenario is the beamformer designs, where the restriction of rank-two beamforming for full-rate transmission results in a multicast design optimization problem that is NP-hard.
The significance of our development lies not only in proposing a rank-two SDR framework for the beamformer design,
but also in generalizing the theoretical analysis of SDR-based beamforming in a non-trivial manner. In particular, we are able to establish for the first time a worst-case performance bound for the NP-hard rank-two beamforming problem.
We should bring readers' attention to the work \cite{wen2012rank,schad2012rank},
wherein the authors independently introduced the same Alamouti-assisted rank-two SDR idea at
about the same time when a preliminary version of this work \cite{wu2012rank} was presented.
What distinguishes our work is that we also provide performance analysis of the resulting scheme.

\subsection{Organization and Notations}

The organization of this paper is as follows. In Section~\ref{sec:formulation} we provide the problem formulation and review the SDR-based multicast beamforming scheme.
The SBF framework is developed and described in Section~\ref{sec:SBF}. Section~\ref{sec:sim} provides the simulation results,
and the paper is concluded in Section~\ref{sec:conclusion}.

Our notation is standard:
$\mathbb{C}^N$ is the set of all complex $N$-dimensional vectors;
$\mathbb{H}^N$ is the set of all $N \times N$ complex Hermitian matrices;
${\bf x} \ge {\bf 0}$ means that ${\bf x}$ is elementwise non-negative;
${\bf X} \succeq {\bf 0}$ means that ${\bf X}$ is positive semidefinite;
$\| \cdot \|$ is the vector Euclidean norm;
${\rm Tr}( {\bf X})$, ${\rm rank}({\bf X})$, ${\rm \lambda}_{\rm max}({\bf X})$, and ${\rm \lambda}_{\rm min}^+({\bf X})$
stand for the trace, rank, the largest eigenvalue, and the smallest non-zero eigenvalue of ${\bf X}$, resp.;
${\bf 0}$ and  ${\bf 1}$ are the all-zero and all-one vectors, resp.;
${\bf e}_i$ is a unit vector with the nonzero element at the $i$th entry;
${\bf I}_r$ denotes the $r$-by-$r$ identity matrix;
$\mathbb{E}[ \cdot ]$ is statistical expectation;
$\mathcal{CN}({\bf 0},{\bf W})$ (resp. $\mathcal{N}({\bf 0},{\bf W})$) is used to denote the circularly symmetric complex Gaussian distribution (resp. the real Gaussian distribution) with mean vector ${\bf 0}$ and covariance matrix ${\bf W}$; and
${\bm X} \sim {\bm Y}$ means that the random variables ${\bm X}$ and ${\bm Y}$ have the same distribution.

\section{Problem Formulation and Background Review}
\label{sec:formulation} \label{sec:fixed_BF}

This section describes the physical-layer multicasting problem formulation and gives a review of multicast beamforming.

We consider a standard multicast scenario~\cite{Conferece:CapacityLimitsLuo06}
where a base station
transmits a common message to $M$ users under slow channel fading.
To be specific, the base station is equipped with $N$ transmit antennas, while the users a single antenna.
The channel of each user is assumed to be frequency flat and slow faded in the sense that its coherence time is larger than the data frame or packet transmission period.
Under this setting,
the signal model for one data frame transmission can be described by
\begin{equation} \label{eq:model}
y_{i}(t)= {\bf h}_{i}^{H} {\bf x}(t)+n_{i}(t), \quad t=1,2,\ldots,T,
\end{equation}
where
$y_i(t)$ is the received signal of user $i$ at time $t$ (or $t$th channel use),
$T$ is the data frame length,
which is assumed to be large,
${\mathbf{h}_{i}} \in \mathbb{C}^N$ is the channel from the base station to user $i$,
$\mathbf{x(}t\mathbf{)}\in \mathbb{C}^{N}$ denotes the multi-antenna transmit signal, and
$n_{i}(t) \sim \mathcal{CN}(0,1)$ is zero mean unit variance complex Gaussian noise.  We denote the transmit covariance by $\bm{\Sigma} = \mathbb{E}[ {\bf x}(t) {\bf x}^H(t) ]$.

The subject of interest is to provide good multicast rate performance for each frame transmission, assuming knowledge of ${\bf h}_1,\ldots,{\bf h}_M$, or channel state information at the transmitter (CSIT).
From an information theoretic perspective, it is known that the multicast capacity under model \eqref{eq:model} and in the presence of CSIT is given by
\begin{equation} \label{eq:MC}
\begin{aligned}
C_{\sf MC}(P) =
    \max_{ \bm{\Sigma} \in \mathbb{H}^N } & ~
    \min_{i=1,\ldots,M} \log( 1 + {\bf h}_i^H \bm{\Sigma} {\bf h}_i ) \\
    {\rm s.t.} & ~ \bm{\Sigma} \succeq {\bf 0},  ~ {\rm Tr}(\bm{\Sigma}) \leq P,
\end{aligned}
\end{equation}
where $P$ is the maximum allowable transmit power and $\log(.)$ is natural logarithm (and thus $C_{\sf MC}(P)$ is in units of nats/s/Hz)~\cite{Conferece:CapacityLimitsLuo06}.
Note that we do not assume any physical-layer transmit structure on ${\bf x}(t)$ at this point.
By the change of variable $\bm{\Sigma} = P {\bf W}$,
we can rewrite~\eqref{eq:MC} as
$$
C_{\sf MC}(P) = \log( 1 + \rho_{\rm min} P ),
$$
where
\begin{equation} \label{eq:rho}
\rho_i = {\rm Tr}( {\bf W}^\star {\bf h}_i {\bf h}_i^H ), \qquad \rho_{\rm min} = \min_{i=1,\ldots,M} \rho_i,
\end{equation}
\begin{equation*}
\begin{aligned}
\text{(MC)} \quad \qquad {\bf W}^\star = \arg \max_{ {\bf W} \in \mathbb{H}^N } & ~ \min_{i=1,\ldots,M} {\rm Tr}( {\bf W} {\bf h}_i {\bf h}_i^H ) \\
    {\rm s.t.} & ~ {\rm Tr}({\bf W}) \leq 1,  ~ {\bf W} \succeq {\bf 0}.
\end{aligned}
\end{equation*}
In particular, an optimal solution $\bm{\Sigma}^\star$ to \eqref{eq:MC} can be constructed from the optimal solution ${\bf W}^\star$ to (MC) via $\bm{\Sigma}^\star= P {\bf W}^\star$.
Problem (MC) is an SDP, which is convex and polynomial-time solvable~\cite{Jnl:MagzineMaLUO}.  Alternatively, one may employ low-complexity heuristics specially designed for (MC); see, e.g.,~\cite{Jnl:successive_BF_2011}.
%

An important question is how physical-layer schemes should be designed to practically approach the information rate promised by the multicast capacity $C_{\sf MC}(P)$.
From such a realizable transceiver design viewpoint, there seems to have no report on a practical multicast capacity-achieving scheme that has been successfully implemented and demonstrated in physical layer.
Currently, a widely adopted scheme is {\it transmit beamforming},
which is efficiently realizable but generally suboptimal.
In transmit beamforming,
the transmit signal ${\bf x}(t)$ is constrained to take the form
\[ {\bf x}(t) = \sqrt{P} {\bf w} s(t), \]
where ${\bf w} \in \mathbb{C}^N$ is a transmit beamforming vector,
$P$ is again the maximum allowable transmit power, and
$s(t) \in \mathbb{C}$ is a stream of data symbols with unit power (i.e., ${\mathbb E}[ |s(t)|^2 ] =1$).
In beamforming, the received signal in \eqref{eq:model} reduces to a single-input single-output (SISO) model
$y_i(t) = \sqrt{P} {\bf h}_i^H {\bf w} s(t) + n_i(t)$,
and we can characterize the performance by the signal-to-noise ratios (SNRs) of the received symbols, namely,
${\sf SNR}_i= P | {\bf h}_i^H {\bf w} |^2$, where $i=1,\ldots,M$.
Consequently, the multicast beamforming problem can be formulated as
\begin{equation} \label{eq:BF}
\max_{ \| {\bf w} \|^{2} \leq 1 } ~ C_{\sf BF}({\bf w},P),
\end{equation}
where
$$ C_{\sf BF}({\bf w},P) = \min_{i= {1,\ldots,M}} \log ( 1 + P {\bf h}^H_i  {\bf w} {\bf w}^H {\bf h}_i ) $$
represents the multicast achievable rate of a given beamformer ${\bf w}$~\cite{MulticastSidiropoulos06,Conferece:CapacityLimitsLuo06}.
Note that this rate can be practically approached by applying an ideal channel code to $s(t)$ \footnote{The common, tacit, understanding is that Turbo codes or low density parity check codes should provide near-ideal scalar channel coding performance in practice.}.
Now, it is known that Problem~\eqref{eq:BF} is equivalent to the max-min-fair (MMF) problem
\begin{equation*}
\begin{aligned}
\text{(MMF)} \quad \qquad \max_{ {\bf w} \in \mathbb{C}^N, ~ \| {\bf w} \|^{2} \leq 1 }   &  ~ \min_{i= {1,\ldots,M}} | {\bf h}_i^H {\bf w} |^2,
\end{aligned}
\end{equation*}
which is NP-hard in general~\cite{MulticastSidiropoulos06,MulticastLuo07}\footnote{Note that the MMF problem was originally formulated from a QoS perspective~\cite{MulticastSidiropoulos06}, where the aim is to maximize the worst user's QoS under a power constraint $\mathbb{E}[ \| {\bf x}(t) \|^2 ] \leq P$.
The QoS commonly refers to the SNR defined here, although other measures of QoS, such as the long-term average SNR~\cite{Lozano_07}, can also be considered.}.
To circumvent this intractability, an arguably {\it de facto} solution is to apply semidefinite relaxation (SDR) to approximate (MMF).
In the SDR approach, one first substitute ${\bf W}= {\bf w}{\bf w}^H$ into (MMF) and use the equivalence
\[ {\bf W} = {\bf w}{\bf w}^H \qquad \Longleftrightarrow \qquad {\bf W} \succeq {\bf 0} \text{~and~} {\rm rank}({\bf W}) \leq 1 \]
to obtain the following {\it equivalent} formulation of (MMF):
\begin{equation} \label{eq:mmf_rank_sdr}
\begin{aligned}
\max_{ {\bf W} \in \mathbb{H}^N } & ~ \min_{i=1,\ldots,M} \mbox{Tr}( \mathbf{W} \mathbf{h}_{i} \mathbf{h}_{i}^{H} ) \\
{\rm s.t.} & ~ {\rm Tr}( \mathbf{W} ) \leq 1, \quad \mathbf{W} \succeq \mathbf{0}, \quad \mbox{rank}(\mathbf{W}) \leq 1.
\end{aligned}
\end{equation}
The rationale behind such a reformulation is that one can then drop the nonconvex rank constraint in \eqref{eq:mmf_rank_sdr} to obtain a convex relaxation problem, viz.
\begin{equation*} 
\begin{aligned}
\text{(SDR)} \qquad \quad \max_{ \mathbf{W} \in \mathbb{H}^{N} } & ~ \min_{i=1,\ldots,M} {\rm Tr}( \mathbf{W} \mathbf{h}_{i} \mathbf{h}_{i}^{H} ) \\
{\rm s.t.} & ~ {\rm Tr}( \mathbf{W} ) \leq 1, \quad \mathbf{W} \succeq \mathbf{0},
\end{aligned}
\end{equation*}
which is an SDP.
Some rank-one approximation procedure is then used to convert the solution of (SDR) to a rank-one, feasible, solution to (MMF);
see \cite{MulticastSidiropoulos06,inbook:Tsung_Hui_bokcha,Jnl:MagzineMaLUO} for details.
It is interesting to note that (SDR) and (MC) are exactly the same.  Hence, the SDR approach essentially uses the multicast capacity-optimal transmit covariance ${\bf W}^\star$ to find a good rank-one beamforming solution.

Empirically, it has been shown that SDR-based multicast beamforming offers good performance, especially for a small to moderate number of users.
In fact, theoretical results quantifying the extent to which SDR can perform are available, and they are briefly summarized as follows. Let
\begin{equation} \label{eq:worst-SNR}
\overline{\sf SNR}_{\rm min}({\bf W}) = \min_{i=1,\ldots,M} {\rm Tr}( {\bf W} {\bf h}_i {\bf h}_i^H )
\end{equation}
denote the worst-user effective SNR associated with ${\bf W}$, which appears in the objective functions of \eqref{eq:mmf_rank_sdr} and (SDR).
By noting that the optimal solution ${\bf W}^\star$ to (MC) is also optimal for (SDR), we have the following:

\begin{Fact} \label{fact:rank-1} {\quad}
\begin{itemize}
\item[(a)] (\cite{Jnl:Yongwei_rank}) When $M \le 3$, there is a polynomial-time procedure that can generate from ${\bf W}^\star$ an optimal solution $\hat{\bf w}$ to (MMF). Also, $\hat{\bf w} \hat{\bf w}^H$ is a solution to (MC).

\item[(b)] (\cite{MulticastLuo07,MulticastSidiropoulos06})
When $M > 3$, by using a Gaussian randomization procedure (which runs in randomized polynomial time), one can generate from ${\bf W}^\star$ a feasible solution $\hat{\bf w}$ to (MMF) that satisfies
\[
 \overline{\sf SNR}_{\rm min}(\hat{\bf w} \hat{\bf w}^H ) \geq \frac{\overline{\sf SNR}_{\rm min}({\bf W}^\star)}{8M} = \frac{\rho_{\rm min}}{8M}
 \]
with probability at least $1/6$.  In particular, after $L\ge1$ independent runs of the randomization procedure, one can boost this probability to at least $1-(5/6)^L$.
\end{itemize}
\end{Fact}

Fact~\ref{fact:rank-1}(a) states that the generally NP-hard (MMF) is equivalent to the convex, polynomial-time solvable (SDR) when the number of users $M$ is no greater than $3$.\footnote{As an aside, note that for $M = 2$, a closed-form solution to (MMF) can be derived~\cite{Jnl:successive_BF_2011}.}
Thus, in view of the equivalence of (SDR) and (MC), we conclude that {\em transmit beamforming is guaranteed to be a multicast capacity-optimal physical-layer strategy for $M \leq 3$.}
As for Fact~\ref{fact:rank-1}(b), it reflects how the performance of SDR-based beamforming scales with the number of users in a worst case sense.
Specifically, consider the achievable rate gap of SDR-based beamforming relative to the multicast capacity, i.e., $C_{\sf MC}(P) - C_{\sf BF}( \hat{\bf w},P )$.
From the derivations above, one can readily deduce the following bound for $M > 3$:
\begin{align}
C_{\sf MC}(P) - C_{\sf BF}( \hat{\bf w},P )  & \leq \log \left( \frac{1+ \rho_{\rm min} P }{ 1 + \rho_{\rm min} P/(8M)} \right).
\label{eq:g_SDR_BF}
\end{align}
Note that for large $P$, the right-hand side of \eqref{eq:g_SDR_BF} is approximately equal to $\log(8M)$, which implies that SDR-based beamforming may suffer from a rate loss that increases logarithmically with the number of users.  Hence, the beamforming strategy is only effective when there are not too many users.

\section{Multicast Stochastic Beamforming}
\label{sec:SBF}

In view of the above mentioned drawbacks of beamforming, in this section we propose an alternative physical-layer multicasting strategy based on stochastic beamforming.

\subsection{System Model}

Consider the following transmit structure:
\begin{equation} \label{eq:xt}
{\bf x}(t)= \sqrt{P} {\bf w}(t) s(t),  \quad t=1,2,\ldots,T,
\end{equation}
where
${\bf w}(t) \in \mathbb{C}^N$ is a time-varying beamformer weight vector,
and the other notations are the same as those in the beamforming strategy discussed above.
At each time $t$, ${\bf w}(t)$ is randomly generated according to a common distribution $\mathcal{D}$.
To distinguish this random-in-time beamforming endeavor from the conventional beamforming scheme, we will henceforth call the former
{\it stochastic beamforming} (SBF), and the latter {\it fixed beamforming}.  The SBF strategy is motivated by the observation that the transmit covariance of \eqref{eq:xt} is given by
$\mathbb{E}[\mathbf{x}(t)\mathbf{x}^{H}(t)] = P \mathbb{E}[\mathbf{w}(t)\mathbf{w}^{H}(t)]$.  In particular, if we choose $\mathcal{D}$ so that the beamformer covariance and the multicast capacity-optimal transmit covariance are equal, i.e.,
\[ \mathbb{E}_{\mathbf{w}(t) \sim \mathcal{D}}[\mathbf{w}(t)\mathbf{w}^{H}(t)]=  {\bf W}^\star, \]
then the SBF should have a better multicast performance than the fixed beamformer, especially when ${\bf W}^\star$ has high rank.  

Let us now consider the receiver side.  Substituting \eqref{eq:xt} into \eqref{eq:model},
the received SBF signals can be written as
\begin{equation} \label{eq:y_sbf}
y_{i}(t)= \sqrt{P} {\bf h}_{i}^{H} {\bf w}(t) s(t)+n_{i}(t), \qquad t=1,2,\ldots,T.
\end{equation}
As seen in \eqref{eq:y_sbf}, each user has an instantaneous SNR given by ${\sf SNR}_i(t) = P | {\bf h}_{i}^{H} {\bf w}(t)|^2$, which fluctuates in time.
Hence,
we
apply channel coding (presumably ideal) across the symbols $\{ s(t) \}_{t=1}^T$ within the data frame
to ``average out'' the fluctuations caused by SBF.  Interestingly, this receiver approach is the same as how one uses channel coding in fast fading channels to exploit time diversity~\cite{Biglieri98fadingchannels}.
We assume coherent reception, which means that all the users are assumed to know ${\bf w}(t)$ deterministically (as well as ${\bf h}_i(t)$).
This can be made possible by having the transmitter sending the random seed for generating $\mathbf{w}(t)$ and the multicast optimal transmit covariance $\mathbf{W}^{\star }$, either as part of the preamble of the transmitted frame or via a feedback channel.
We should also note that SBF receivers involve simple coherent symbol reception (without inter-symbol interference) and channel decoding,
and hence are as efficient as those of fixed beamforming with channel coding.

The SBF system description is complete.
Now, several natural questions arise:
What distribution $\mathcal{D}$ should we use to generate the random beamformer weights?
How can we characterize the performance of an SBF scheme?
These aspects are considered in the subsequent subsections.

\subsection{SBF Achievable Rate}
\label{sec:SBF_rate}

We employ an achievable rate view to study the SBF strategy.
For notational simplicity, we use the random variable $\bm{w}$ to denote the randomly generated beamformer weight vector ${\bf w}(t)$.
Under the SBF system model in \eqref{eq:y_sbf},
where channel coding is applied across $\{ s(t) \}_{t=1}^T$ with $T$ sufficiently large,
the achievable rate of each user, say, user $i$, can be expressed as
\begin{equation} \label{eq:C_SBF_i}
C_{{\sf SBF},i}(P) = \mathbb{E}_{\bm{w} \sim \mathcal{D}}[\log (1+ P {\bf h}_{i}^{H}{ \bm{w}\bm{w}^{H} } {\bf h}_{i})],
\end{equation}
where $\mathcal{D}$ denotes the (given) distribution for generating $\bm{w}$.
We should mention that the capacity expression in \eqref{eq:C_SBF_i} is deduced in the same spirit as the ergodic capacities for fast fading channels without CSIT, as described or used frequently in the literature; see, e.g., \cite{Biglieri98fadingchannels,Yeung:2006ITbook}.
However, we should emphasize that in this study, it is not the channels ${\bf h}_i$ that are random, but the beamformer $\bm{w}$.
Moreover,
studies in
fast fading channels
have suggested that
the rate \eqref{eq:C_SBF_i} may practically be approached by near-ideal scalar channel codes; see, e.g., \cite[p.~2627]{Biglieri98fadingchannels}.
Based on \eqref{eq:C_SBF_i}, the multicast achievable rate of SBF can be formulated as
\begin{equation} \label{eq:C_SBF}
C_{\sf SBF}(P)=  \min_{i=1,\ldots,M}\mathbb{E}_{\bm{w} \sim \mathcal{D}}[\log (1+ P {\bf h}_{i}^{H}{ \bm{w}\bm{w}^{H} } {\bf h}_{i} )].
\end{equation}
Note that
$\mathcal{D}$ must satisfy $ \mathbb{E}_{ \bm{w} \sim \mathcal{D}} [ \| \bm{w} \|^2 ] \leq 1$, so that the power constraint $\mathbb{E}[ \| {\bf x}(t) \|^2 ] \leq P$ holds.

Before we proceed, let us discuss the key underlying assumption behind the SBF achievable rate metric above---that $T$ should be large.
In practice, the frame length $T$ is constrained by the coherence time of the channels.
As such, the rate metric above is more suitable for slow fading scenarios.
In our simulations, we found that the idea works well when $T$ is the same as that of the coded symbol length for a fixed beamforming channel (or a standard scalar Gaussian channel), which is typically on the order of hundreds in wireless standards.

To facilitate the SBF design and rate analysis, we first derive an alternative expression for $C_{\sf SBF}(P)$.
Set
\begin{equation} \label{eq:xi_i}
\xi_i = \frac{ | {\bf h}_{i}^{H} \bm{w} |^2 }{\rho_i }, \quad i=1,\ldots,M
\end{equation}
(see \eqref{eq:rho} for the definition of $\rho_i$).
Clearly, if $\mathcal{D}$ satisfies the capacity-optimal transmit covariance property $\mathbb{E}_{\bm{w} \sim \mathcal{D}}[ \bm{w} \bm{w}^H ] = {\bf W}^\star$, then $\mathbb{E}[ \xi_i ] = 1$.
Then, we can rewrite \eqref{eq:C_SBF} as
\begin{equation} \label{eq:C_SBF_2}
C_{\sf SBF}(P)=  \min_{i=1,\ldots,M} \mathbb{E}_{\xi_i }[\log (1+ \xi_i \rho_i P )].
\end{equation}
The above SBF rate characterization reveals that the SBF performance depends on the ``fading'' distribution of $\xi_i$.
The following properties can be derived for \eqref{eq:C_SBF_2}:
\begin{Fact} \label{fact:SBF_basic}
Suppose that $\xi_1,\ldots,\xi_M$ are identically distributed.
Let $\xi \sim \xi_i$ for any $i$.
\begin{itemize}
\item[(a)]  The SBF multicast achievable rate \eqref{eq:C_SBF_2} can be simplified to
$C_{\sf SBF}(P) = \mathbb{E}_{\xi }[\log (1+ \xi \rho_{\rm min} P )]$.
\item[(b)] Suppose, in addition, that $\mathbb{E}[ \xi_i ] = 1$.
Then, the function $g_{\sf SBF}:{\mathbb R}_+\rightarrow{\mathbb R}_+$, where
$g_{\sf SBF}(P)= C_{\sf MC}(P) - C_{\sf SBF}(P)$, is
nondecreasing
in $P \geq 0$.
\end{itemize}
\end{Fact}
Fact~\ref{fact:SBF_basic}(a) is simply a consequence of the monotonicity of the log function.  For a proof of Fact~\ref{fact:SBF_basic}(b), see Appendix~\ref{sec:A}.

\subsection{The Gaussian SBF Scheme}
\label{sec:SBF_Gauss}

Let us now turn our attention to the choice of the beamformer distribution $\mathcal{D}$.
The most desirable choice of $\mathcal{D}$ would be that of maximizing the multicast achievable rate under the power constraint.
However, this may be too difficult to solve analytically.
Hence, we seek simple, easy-to-generate, beamformer randomizations that can yield provably good multicast rate performance.


A simple way to generate $\bm{w}$ is to use the circularly symmetric complex Gaussian distribution:
\begin{equation} \label{eq:w_gauss}
   \bm{w} \sim \mathcal{CN}( {\bf 0}, {\bf W}^\star ).
\end{equation}
We will call the resulting SBF scheme the {\it Gaussian SBF scheme} .
Gaussian SBF aims at using a simple beamformer generation to satisfy the optimal transmit covariance property $\mathbb{E}[ \bm{w} \bm{w}^H ] = {\bf W}^\star$.
It can be analytically shown that even such a simple beamformer randomization possesses desirable multicast achievable rate properties.
From \eqref{eq:xi_i}, we see that for Gaussian SBF, every $\xi_i$ follows an exponential distribution with mean $\mathbb{E}[ \xi_i ] =1$.
Therefore, the premises of Fact~\ref{fact:SBF_basic} are satisfied,
and by Fact~\ref{fact:SBF_basic}(a) we can express the Gaussian SBF achievable rate as
\begin{equation} \label{eq:SBF_Gaus0}
C_{\sf SBF}^{\sf Gauss}(P) = \int_0^\infty \log( 1 + t \rho_{\rm min} P ) e^{-t} dt.
\end{equation}
As it turns out, the expression in \eqref{eq:SBF_Gaus0} is identical to that for the ergodic capacity of a scalar Rayleigh channel, which is known to admit the explicit expression
\begin{equation} \label{eq:SBF_Gaus}
C_{\sf SBF}^{\sf Gauss}(P) = e^{1/(\rho_{\rm min} P)} E_1( 1/(\rho_{\rm min} P) ),
\end{equation}
where $E_1(x) = \int_1^\infty t^{-1} e^{-xt} dt$, $x \geq 0$, is the exponential integral of the first order~\cite{Alouini1999}.
Now, we are interested in extracting insight from the explicit rate expression \eqref{eq:SBF_Gaus}---how far away is \eqref{eq:SBF_Gaus} from the
multicast capacity $C_{\sf MC}(P)$?  Towards that end, consider the achievable rate gap
\[ g_{\sf SBF}^{\sf Gauss}(P) = C_{\sf MC}(P) - C_{\sf SBF}^{\sf Gauss}(P). \]
We then have the following result:
\begin{Theorem} \label{thm:SBF_Gaus}
The achievable rate gap of the Gaussian SBF scheme satisfies
\[ g_{\sf SBF}^{\sf Gauss}(P) \leq \gamma = 0.5772 \qquad \text{for all $P \geq 0$.} \]
Moreover, the bound is tight when $P \rightarrow \infty$.
\end{Theorem}

\noindent {\it Proof:} \
By Fact~\ref{fact:SBF_basic}(b), $g_{\sf SBF}^{\sf Gauss}(P)$ is nondecreasing in $P \geq 0$.
Moreover, it can be shown that $\lim_{P \rightarrow \infty} g_{\sf SBF}^{\sf Gauss}(P) = \gamma$; see Section 1.1 of the companion technical report~\cite{CompTechRep}.
Hence, we conclude that $g_{\sf SBF}^{\sf Gauss}(P) \leq \gamma$ for all $P \geq 0$.
\hfill $\blacksquare$

The implication of Theorem~\ref{thm:SBF_Gaus} is meaningful---the Gaussian SBF rate $C_{\sf SBF}^{\sf Gauss}(P)$ is at most $0.8314$~bits/s/Hz ($\gamma/\log(2)= 0.8314$) away from the multicast capacity $C_{\sf MC}(P)$;
otherwise it has the same scaling as the multicast capacity, {\it irrespective of the number of users}.
This is unlike the SDR-based fixed beamforming scheme reviewed in Section~\ref{sec:fixed_BF}, where the rate gap may increase with the number of users;
cf.~\eqref{eq:g_SDR_BF}.

\subsection{The Elliptic SBF Scheme}
\label{sec:SBF_elli}

As shown in the previous subsection, even with just a simple Gaussian SBF scheme, we can achieve a rate that is within less than 1 bit/s/Hz of the
multicast capacity.  From a practical viewpoint, however, the Gaussian SBF scheme has a drawback---its instantaneous beamformer power, which is given by $P \|{\bf w}(t)\|^2$, can have a large spread.  Indeed, since $\|\bm{w}\|^2$ is a chi-square random variable, the instantaneous power can in principle take any non-negative values.
Hence, while Gaussian SBF is interesting from a fundamental viewpoint, where a theoretically provable rate gap of less than one bit w.r.t. the multicast capacity can be established, it may not be desirable for practical implementation.
To remedy this, we consider an alternative SBF scheme, in which the beamformer weight is generated by
\begin{equation} \label{eq:w_ellip}
\bm{w} = \frac{ {\bf L}^H \bm{\alpha} }{ \| \bm{\alpha} \| / \sqrt{r} }, \quad \bm{\alpha} \sim \mathcal{CN}( {\bf 0}, {\bf I}_r ),
\end{equation}
where $r=\mbox{rank}({\bf W}^\star)$ and ${\bf L} \in {\mathbb C}^{r\times N}$ is a square root decomposition of ${\bf W}^\star$, i.e., ${\bf L}^H{\bf L} = {\bf W}^\star$.  Note that~\eqref{eq:w_ellip} is simply a Gaussian SBF normalized by the factor $\| \bm{\alpha} \| /\sqrt{r}$; cf.~\eqref{eq:w_gauss}.  Intuitively, such a normalization serves to limit the instantaneous beamformer power.
More precisely, since ${\rm Tr}({\bf W}^\star) \le 1$, by the Courant-Fischer min-max theorem, we have $\| \bm{w} \|^2 \in [ r \lambda_{\rm min}^+({\bf W}^\star), r  \lambda_{\rm max}({\bf W}^\star) ]$ with probability 1.  As it turns out, the random vector $\bm{w}$ also satisfies the capacity-optimal transmit covariance property:
\begin{Fact} \cite{Bok:Multivariate90} \label{fact:ellip}
The random vector in \eqref{eq:w_ellip} follows an elliptic symmetric distribution with covariance matrix $\mathbb{E}[ \bm{w} \bm{w}^H ] = {\bf W}^\star$.
\end{Fact}
Motivated by Fact \ref{fact:ellip}, we shall call the resulting SBF scheme the {\it elliptic SBF scheme}.  Now, just as in the case of the Gaussian SBF scheme, we are interested in determining the achievable rate of the elliptic SBF scheme.  Towards that end, consider the non-negative random variables
\begin{equation} \label{eq:ellip_xi}
   \xi_i = \frac{ | {\bf h}_{i}^{H} {\bf L}^H \bm{\alpha} |^2 }{ \rho_i \| \bm{\alpha} \|^2/r }, \quad i=1,\ldots,M;
\end{equation}
see \eqref{eq:xi_i}.  Naturally, we would like to use Fact \ref{fact:SBF_basic} to characterize the elliptic SBF rate.  However, this entails understanding the distribution of $\xi_i$.  Fortunately, as we shall see shortly, the distribution of $\xi_i$ admits a simple closed form expression.  We begin with the following lemma, which generalizes~\cite[Lemma 1]{ZS11} and whose proof can be found in Appendix \ref{sec:B}:
\begin{Lemma} \label{lem:ellip_dist_gen}
   Let ${\bf u} \in \mathbb{C}^r$ be a fixed vector and $\bm{\alpha}_1,\ldots,\bm{\alpha}_l \sim \mathcal{CN}( {\bf 0}, {\bf I}_r )$ be independent random vectors.  Then, the CDF of the non-negative random variable
$$ \eta({\bf u}) = \sum_{i=1}^l |{\bf u}^H\bm{\alpha}_i|^2  \Bigg/ \sum_{i=1}^l \|\bm{\alpha}_i\|^2 $$
is given by
$$
  \Pr( \eta({\bf u}) \le t) = \left\{
  \begin{array}{l@{\quad}l}
     0 & \mbox{for } t < 0, \\
     \noalign{\smallskip}
     1-Q({\bf u},t) & \mbox{for } 0 \le t \le \|{\bf u}\|^2, \\
     \noalign{\smallskip}
     1 & \mbox{for } t > \|{\bf u}\|^2,
  \end{array}
  \right.
$$
where
$$
Q({\bf u},t) =  1 - \left( \frac{t}{\|{\bf u}\|^2} \right)^{lr-1} \sum_{j=l(r-1)}^{lr-1} {{lr-1} \choose j} \left( \frac{ \|{\bf u}\|^2 - t}{t} \right)^j.
$$
\end{Lemma}
From~\eqref{eq:ellip_xi}, we see that if we take ${\bf u} = \sqrt{r/\rho_i} {\bf L}{\bf h}_i$ and $l=1$ in Lemma \ref{lem:ellip_dist_gen}, then $\xi_i = \eta({\bf u})$.  In particular, upon differentiating the corresponding CDF w.r.t.~$t$ and observing that $\| {\bf u} \|^2 = r$, we obtain the following:
\begin{Prop} \label{prop:ellip_dist}
  Consider the elliptic SBF scheme.  The PDF of $\xi_i$, where $i=1,\ldots,M$, is given by
  \begin{equation} \label{eq:ellip_dist_fn}
  p_{\xi_i}(t) = \left( 1 - \frac{1}{r} \right) \left( 1 - \frac{t}{r} \right)^{r-2} \quad\mbox{for } 0 \le t \le r,
  \end{equation}
where $r={\rm rank}( {\bf W}^\star)$.
\end{Prop}

Proposition \ref{prop:ellip_dist} has two important implications.  First, it shows that the random variables $\xi_1,\ldots,\xi_M$ are identically distributed, and hence by \eqref{eq:ellip_dist_fn} and Fact \ref{fact:SBF_basic}(a) the elliptic SBF rate can be readily computed via
$$ C_{\sf SBF}^{\sf Ellip}(P) = \left( 1 - \frac{1}{r} \right) \int_0^r \log( 1 + t \rho_{\rm min} P ) \left( 1 - \frac{t}{r} \right)^{r-2} dt. $$
Secondly, we have $\mathbb{E}[ \xi_i ] = 1$ by Fact \ref{fact:ellip}.  Hence, by Fact \ref{fact:SBF_basic}(b), the achievable rate gap of the elliptic SBF scheme, which is given by
$$ g_{\sf SBF}^{\sf Ellip}(P) = C_{\sf MC}(P) - C_{\sf SBF}^{\sf Ellip}(P), $$
is nondecreasing in $P\ge0$.

To further understand the behavior of $g_{\sf SBF}^{\sf Ellip}(P)$, let us first derive an explicit formula for $C_{\sf SBF}^{\sf Ellip}(P)$.
\begin{Prop} \label{prop:ellip-rate}
For any $P > 0$,
\ifconfver
\begin{align*}
C_{\sf SBF}^{\sf Ellip}(P) &= \left( 1 + \frac{1}{r\rho_{\rm min}P} \right)^{r-1} \left[ \log(1+r\rho_{\rm min}P) - \sum_{k=1}^{r-1}\frac{1}{k} \right. \\
&\quad \qquad\qquad\quad \left. - \sum_{k=1}^{r-1} {r-1 \choose k} \frac{(-1)^k}{k(1+r\rho_{\rm min}P)^k} \right].
\end{align*}
\else
\begin{align*}
C_{\sf SBF}^{\sf Ellip}(P) &= \left( 1 + \frac{1}{r\rho_{\rm min}P} \right)^{r-1} \left[ \log(1+r\rho_{\rm min}P) - \sum_{k=1}^{r-1}\frac{1}{k}  - \sum_{k=1}^{r-1} {r-1 \choose k} \frac{(-1)^k}{k(1+r\rho_{\rm min}P)^k} \right].
\end{align*}
\fi
\end{Prop}
The proof of Proposition \ref{prop:ellip-rate} can be found in Section 1.2.1 of the companion technical report~\cite{CompTechRep}.  Armed with this formula, we can establish the following result:
\begin{Theorem} \label{thm:SBF_Ellip}
The achievable rate gap of the elliptic SBF scheme satisfies
\[ g_{\sf SBF}^{\sf Ellip}(P) \leq \sum_{k=1}^{r-1} \frac{1}{k} - \log(r) \qquad \text{for all $P \geq 0$.} \]
Moreover, the bound is tight when $P \rightarrow \infty$.
\end{Theorem}

\noindent {\it Proof:} \
We have already shown that $g_{\sf SBF}^{\sf Ellip}(P)$ is nondecreasing in $P \geq 0$.  Moreover, it can be shown that $\lim_{P \rightarrow \infty} g_{\sf SBF}^{\sf Ellip}(P) = \sum_{k=1}^{r-1} \frac{1}{k} - \log(r)$; see Section 1.2.2 of the companion technical report~\cite{CompTechRep}.  Hence, we conclude that $g_{\sf SBF}^{\sf Ellip}(P) \leq \gamma$ for all $P \geq 0$.
\hfill $\blacksquare$

Since the function $r \mapsto \sum_{k=1}^{r-1} \frac{1}{k} - \log(r)$ is nondecreasing and tends to $\gamma$ as $r \rightarrow \infty$ (see, e.g., \cite[Formula 0.131]{bk:Gradshteyn}), an important corollary of Theorem \ref{thm:SBF_Ellip} is that the worst-case rate gap of the elliptic SBF scheme is no worse than that of the Gaussian SBF scheme.  For comparison, we compute the worst-case rate gap of the elliptic SBF scheme for various values of $r$ and summarize the results in Table \ref{tab:1}.

\begin{table}[h]

\caption{\protect{The worst-case rate gap of the elliptic SBF scheme}}
\vspace*{-.5\baselineskip}
\renewcommand{\arraystretch}{1.3} \centering
\begin{tabular}{|c|c|c|c|c|c|}
\hline $r$ &~~~$1$~~~~& $2$& $3$& ~$\ldots$~~& $\infty$
\\\hline rate gap in nats & $0$ & $0.3069$ &$0.4014$& $\ldots$&$0.5772$
\\\hline rate gap in bits & $0$ & $0.4428$ &$0.5791$& $\ldots$&$0.8327$\\\hline
\end{tabular}
\label{tab:1}
\vspace*{-1.5\baselineskip}
\end{table}


\subsection{The Bingham SBF Scheme}
\label{sec:SBF_bing}

In the previous subsection,
we have illustrated that a proper normalization of the Gaussian beamformer randomization not only helps to limit the instantaneous beamformer power spread effects,
but also improves the multicast achievable rate.
Now, let us consider another beamformer randomization
\begin{equation} \label{eq:w_bing}
\bm{w} = \frac{ {\bf L}^H \bm{\alpha} }{ \| {\bf L}^H \bm{\alpha} \| }, \quad \bm{\alpha} \sim \mathcal{CN}( {\bf 0}, {\bf I}_r ).
\end{equation}
The motivation behind \eqref{eq:w_bing} is straightforward---we want $\| \bm{w} \|^2 =1$, or in other words, zero instantaneous beamformer power spread.
Curiously, the kind of randomization in \eqref{eq:w_bing} has been studied in the statistics literature---it is known that $\bm{w}$ follows the {\it Bingham distribution}~\cite{Jnl:complexbingham94}.
For that reason, we will call the resulting SBF scheme the {\it Bingham SBF scheme}.

%

Unlike the previous two SBF schemes, Bingham SBF may not satisfy the capacity-optimal transmit covariance property $\mathbb{E}[ \bm{w} \bm{w}^H ] = {\bf W}^\star$.
Moreover, the achievable rate analysis of Bingham SBF is different from that of Gaussian and elliptic SBF---a key component of the latter is to derive the distribution of $\xi_i$ in \eqref{eq:xi_i}, and this appears to be hard for Bingham SBF.
We therefore resort to a different analysis approach.
Consider the following proposition, whose proof can be found in Appendix \ref{sec:C}:

\begin{Prop} \label{prop:bing_rate}
For the Bingham SBF scheme, the rate of user $i$ can be expressed as
\begin{align}
C_{{\sf SBF},i}^{\sf Bing}(P) &=
\mathbb{E}_{\bm{w}}[ \log( 1 + P | {\bf h}_i^H {\bm w} |^2 ) ] \nonumber \\
&= \log( 1 + \rho_i P ) +
    \varphi\left( \frac{\bm{\mu}_i}{ \bm{\mu}_i^T {\bf 1} } \right)
    - \varphi\left( \bm{\lambda} \right).  \label{eq:bing_rate}
\end{align}
Here, $\varphi: \mathbb{R}^r \rightarrow \mathbb{R}$ is given by
\begin{equation} \label{eq:varphi}
\varphi\left( {\bf d} \right) = \mathbb{E}_{\bm{\zeta}}\left[ \log\left( \sum_{k=1}^r d_k \zeta_k \right)   \right],
\end{equation}
where $\bm{\zeta}$ is a random vector with
independent and identical (i.i.d.) unit-mean exponentially distributed components,
$\bm{\lambda}= (\lambda_1,\ldots,\lambda_r)$ contains the positive eigenvalues of ${\bf W}^\star$, and
$\bm{\mu}_i= (\mu_{i,1},\ldots,\mu_{i,r})$ contains the eigenvalues of
${\bf A}_i = {\bf L} ( {\bf I}_N + P {\bf h}_i {\bf h}_i^H ) {\bf L}^H$.
\end{Prop}

As it turns out, one can derive an explicit expression for $\varphi( {\bf d} )$.
\begin{Prop} \label{fact:sumofexp}
Let $\varphi$ be as in \eqref{eq:varphi}. Organize $\mathbf{d}$ as
\[
\mathbf{d} = (\underbrace{\tilde d_1,\ldots,\tilde d_1}_{r_1},\underbrace{\tilde d_2,\ldots,\tilde d_2}_{r_2},\ldots,\underbrace{\tilde d_c,\ldots,\tilde d_c}_{r_c}),
\]
where $c, r_1,\ldots,r_c$ are such that $\sum_{i=1}^{c}r_i=r$, and $\tilde d_i \neq \tilde d_j$ for all $i \neq j$.
Then, we have
\ifconfver
\begin{align*}
& \varphi\left( {\bf d} \right) \nonumber \\
&= {\prod_{n=1}^{c}\frac{1}{\tilde d_n^{r_n}}
\sum_{k=1}^{c}\sum_{m=1}^{r_k}\frac{\Psi_{k,m,\mathbf{r}}}{(r_k-m)!}(-1)^{(r_k-m)}\theta({\tilde d}_k, r_k-m),} 
\end{align*}
where ${\bf{r}}=(r_1, r_2,\ldots,r_c)$, $\mathbf{i}=(i_1,i_2,\ldots,i_c)$,
\begin{align*}
& \theta({\tilde d}_k, r_k-m) = \tilde d_k^{(r_k-m+1)} \times (r_k-m)! \,\times \nonumber\\
& \qquad\qquad\qquad\qquad \left(\sum_{ i=1}^{r_k-m}\frac{1}{i}+\log(\tilde{d}_k)-\gamma \right), \\ 
& \Psi_{k,m,\mathbf{r}} = (-1)^{(r_k-1)} \times \\ 
& \qquad\qquad\quad \sum_{ {\bf i} \in \Omega_{k,m}} \prod_{j \neq k} {i_j+r_j-1 \choose i_j}
\Big(\frac{1}{\tilde d_j}-\frac{1}{\tilde d_k} \Big)^{-(i_j+r_j)}, \\ 
& \Omega_{k,m} = \left\{ {\bf i} \in {\mathbb{Z}^{c}} : \sum_{j=1}^{c}i_j=m-1, \, i_k=0,\, i_j\ge0 \,\,\,\forall j \right\}. 
\end{align*}
\else
\begin{align*}
\varphi\left( {\bf d} \right) & = {\prod_{n=1}^{c}\frac{1}{\tilde d_n^{r_n}}
\sum_{k=1}^{c}\sum_{m=1}^{r_k}\frac{\Psi_{k,m,\mathbf{r}}}{(r_k-m)!}(-1)^{(r_k-m)}\theta({\tilde d}_k, r_k-m),} 
\end{align*}
where ${\bf{r}}=(r_1, r_2,\ldots,r_c)$, $\mathbf{i}=(i_1,i_2,\ldots,i_c)$,
\begin{align*}
\theta({\tilde d}_k, r_k-m) & = \tilde d_k^{(r_k-m+1)} \times (r_k-m)! \,\times \left(\sum_{ i=1}^{r_k-m}\frac{1}{i}+\log(\tilde{d}_k)-\gamma \right), \\ 
 \Psi_{k,m,\mathbf{r}} & = (-1)^{(r_k-1)} \times
 \sum_{ {\bf i} \in \Omega_{k,m}} \prod_{j \neq k} {i_j+r_j-1 \choose i_j}
\Big(\frac{1}{\tilde d_j}-\frac{1}{\tilde d_k} \Big)^{-(i_j+r_j)}, \\ 
 \Omega_{k,m} & = \left\{ {\bf i} \in {\mathbb{Z}^{c}} : \sum_{j=1}^{c}i_j=m-1, \, i_k=0,\, i_j\ge0 \,\,\,\forall j \right\}. 
\end{align*}
\fi
\end{Prop}
The proof of Proposition~\ref{fact:sumofexp} can be found in Section 2 of the companion technical report~\cite{CompTechRep}.
The idea behind the proof of Proposition~\ref{prop:bing_rate} is somewhat similar to that in \cite[Theorem~1]{Bjornson2010},
where the authors there dealt with a different scenario (unicast).
While Proposition~\ref{fact:sumofexp} gives an explicit expression for \eqref{eq:varphi},
which in turn provides a way of computing the Bingham SBF achievable rate efficiently (in contrast with Monte Carlo simulations),
it is too complicated for the purpose of extracting insights.
This difficulty motivates us to turn to the stochastic majorization technique for Bingham SBF rate gap characterization:

\begin{Fact} \label{fact:varphi}
Consider 
$\varphi\left( {\bf d} \right) = \mathbb{E}_{\bm{\zeta}}\left[ \log\left( \sum_{k=1}^n d_k \zeta_k \right)   \right]$,
where $\bm{\zeta}$ is a random vector with arbitrary i.i.d. components.
\begin{itemize}
\item[(a)] (\cite[Theorem~2.15, Example~2.2]{Jorswieck_now_book2007}) For any ${\bf d}= (d_1,\ldots,d_n) \ge {\bf 0}$ with $\sum_{k=1}^n d_k = 1$,
$$ 
\varphi( {\bf e}_1 ) \leq \varphi({\bf d}) \leq \varphi\left( \tfrac{1}{n} {\bf 1}  \right).
$$ 
\item[(b)] (\cite{Jnl:Moser2003})
Suppose that every $\xi_i$ follows a unit-mean exponential distribution. Then, we have
\[ \varphi\left( \tfrac{1}{n} {\bf 1} \right) = \sum_{k=1}^{n-1} \frac{1}{k} - \log(n) - \gamma. \]
\end{itemize}
\end{Fact}

Applying Fact~\ref{fact:varphi} to \eqref{eq:bing_rate}, we obtain
\begin{align*}
\mathbb{E}_{\bm{w}}[ \log( 1 + P | {\bf h}_i^H {\bm w} |^2 ) ]
    & \geq \log( 1 + \rho_i P ) + \varphi( {\bf e}_1) - \varphi\left( \tfrac{1}{r} {\bf 1} \right) \\ 
    & \geq \log( 1 + \rho_i P ) + \log(r) - \sum_{k=1}^{r-1} \frac{1}{k}, 
\end{align*}
where the first inequality follows from Fact~\ref{fact:varphi}(a) and the observation that $\sum_{k=1}^r \lambda_k = {\rm Tr}( {\bf W}^\star ) = 1$ (this is implied by the structure of (MC)), and the second inequality is
due to Fact~\ref{fact:varphi}(b).  The derivations above show that user-$i$'s Bingham rate is lower bounded by $\log( 1 + \rho_i P ) + \log(r) - \sum_{k=1}^{r-1} \frac{1}{k}$, which lead us to a neat conclusion:

\begin{Theorem} \label{thm:SBF_Bing}
The achievable rate gap
$g_{\sf SBF}^{\sf Bing}(P)= C_{\sf MC}(P) - C_{\sf SBF}^{\sf Bing}(P)$
of the Bingham SBF scheme satisfies
\[ g_{\sf SBF}^{\sf Bing}(P) \leq \sum_{k=1}^{r-1} \frac{1}{k} - \log(r) \qquad \text{for all $P \geq 0$.} \]
\end{Theorem}

Surprisingly, the worst-case rate gap of the Bingham SBF scheme as proven above is exactly the same as that of the elliptic SBF scheme (cf. Theorem~\ref{thm:SBF_Ellip}).
It follows that the worst-case rate gap of the Bingham SBF scheme is also no worse than that of the Gaussian SBF scheme.

\subsection{Summary of the SBF Schemes}

We now summarize the characteristics of our proposed SBF schemes in Table~\ref{tab:SBF_summary}.
It can be seen that all three schemes exhibit a multicast achievable rate gap that is no worse than $0.8314$~bits/s/Hz, {\it irrespective of any factors such as the number of users}.
In fact, the elliptic and Bingham SBF schemes can perform better than $0.8314$~bits/s/Hz, depending on the transmit covariance rank $r= {\rm rank}({\bf W}^\star)$;
see Table~\ref{tab:1}.
In terms of the instantaneous beamformer power spread effects,
the Gaussian SBF scheme is, by nature, the worst.
The elliptic SBF scheme is better than the Gaussian SBF scheme, limiting the instantaneous beamformer power to within $[ r \lambda_{\rm min}^+({\bf W}^\star), r  \lambda_{\rm max}({\bf W}^\star)]$.
The Bingham SBF scheme has zero instantaneous beamformer power spread.  On the other hand, the Gaussian and elliptic SBF schemes achieve the multicast capacity-optimal transmit covariance $\mathbb{E}[ \bm{w} \bm{w}^H ] = {\bf W}^\star$,
while the Bingham SBF scheme may not.

\ifconfver
    \begin{table*}[th]
\else
    \begin{table}[h]
\fi
\caption{Summary of the SBF schemes}
\label{tab:SBF_summary}
\begin{center}
\linespread{1.25} \rm \footnotesize
\begin{tabular}{c||c|c|c|c}
scheme & generation &
\begin{minipage}{0.15\textwidth}
\begin{center}
has MC-opt. covariance \\ $\mathbb{E}[ \bm{w}\bm{w}^H ]=\bf{W}^{\star}$?
\end{center}
\end{minipage}
&
\begin{minipage}{0.2\textwidth}
\begin{center}
instantaneous beamformer \\ power spread
\end{center}
\end{minipage}
&
\begin{minipage}{0.18\textwidth}
\begin{center}
worst-case rate gap \\ upper bound
\end{center}
\end{minipage} \\ \hline\hline
Gaussian &
$\bm{w} \sim \mathcal{CN}( {\bf 0}, {\bf W}^\star )$ &
yes  & large & $0.8314$ bits/s/Hz \\ \hline
elliptic &
\begin{minipage}{0.20\textwidth}
\[ \bm{w} = \frac{ {\bf L}^H \bm{\alpha} }{ \| \bm{\alpha} \| / \sqrt{r} },\]
where
$\bm{\alpha} \sim \mathcal{CN}( {\bf 0}, {\bf I}_r )$;
${\bf L} \in {\mathbb C}^{r\times N}$ is a square root factor of ${\bf W}^\star$;
$r= {\rm rank}({\bf W}^\star)$
\end{minipage}
& yes &
\begin{minipage}{0.285\textwidth}
better than Gaussian;
\[ \| \bm{w} \|^2 \in [ r \lambda_{\rm min}^+({\bf W}^\star), r  \lambda_{\rm max}({\bf W}^\star)] \]
with probability $1$
\end{minipage}
&
\begin{minipage}{0.18\textwidth}
\[
\begin{array}{rl}
& 
\displaystyle \frac{ \sum_{k=1}^{r-1} \frac{1}{k} - \log(r) }{\log(2)} \\
& \leq \text{$0.8314$ bits/s/Hz;}
\end{array}
\]
optimal when $r=1$
\end{minipage}  \\ \hline
Bingham  &
\begin{minipage}{0.20\textwidth}
\[ \bm{w} = \frac{ {\bf L}^H \bm{\alpha} }{ \| {\bf L}^H \bm{\alpha} \| },\]
where
$\bm{\alpha} \sim \mathcal{CN}( {\bf 0}, {\bf I}_r )$.
\end{minipage} &
no & zero; $\| \bm{w} \|^2 = 1$
 & same as elliptic
\end{tabular}
\end{center}
\ifconfver
    \end{table*}
\else
    \end{table}
\fi

\section{Multicast Beamformed Alamouti Space-Time Coding}
\label{sec:BF_alam}

In this section, we describe our second physical-layer multicasting strategy---transmit beamformed Alamouti space-time coding.
Compared to SBF, which uses time randomizations to enable rank-$r$ transmit covariance structures, the beamformed Alamouti strategy adopts a rank-two transmit covariance structure in a fixed or deterministic way. This will motivate a rank-two generalization of SDR.

\subsection{System Model}
\label{sec:alam_model}

We describe the system model for (fixed) beamformed Alamouti space-time coding.
Like the beamforming case, we aim at transmitting a stream of unit-power data symbols, denoted by $s(t)$.
The data symbol stream $s(t)$ is parsed into blocks via
 ${\bm s}(n)= [~ s(2n) ~ s(2n+1) ~]^T$.
In block $n$, we transmit ${\bm s}(n)$ by a transmit beamformed Alamouti space-time code:
\begin{equation} \label{eq:X_codeblk}
{\bm X}(n) = [~ {\bf x}(2n) ~ {\bf x}(2n+1) ~] = \sqrt{P} {\bf B} \mathbf{C}( {\bm s}(n) ).
\end{equation}
Here, ${\bf B} \in \mathbb{C}^{N \times 2}$ is a transmit beamforming matrix
and $\mathbf{C}: \mathbb{C}^2 \rightarrow \mathbb{C}^{2 \times 2}$ is the Alamouti space-time block code, i.e.,
\[ {\bf C}({\bm s}) = \begin{bmatrix} s_1 & s_2 \\ -s_2^* & s_1^* \end{bmatrix}. \]
From the basic model in \eqref{eq:model},
we have
\begin{equation} \label{eq:model_alam}
{\bm y}_i(n) = [~ y_i(2n) ~ y_i(2n+1) ~]= \sqrt{P} {\bf h}_i^H \mathbf{B} \mathbf{C}( {\bm s}(n) ) + {\bm n}_i(n),
\end{equation}
where ${\bm n}_i(n) = [~ n_i(2n) ~ n_i(2n+1) ~]$.
Using a key property introduced by the special structure of the Alamouti code (see, e.g., \cite{larsson2008space}),
Eq.~\eqref{eq:model_alam} can be turned into an equivalent SISO model,
where each symbol can be independently detected
and user $i$'s SNR of the received symbols can be characterized by
${\sf SNR}_i = P {\bf h}_i^H \mathbf{B}  \mathbf{B}^H {\bf h}_i.$
Hence, for the beamformed Alamouti strategy, we can formulate the following achievable rate problem:
\begin{equation} \label{eq:mmf_alam}
\begin{aligned}
 \max_{ {\bf B} \in \mathbb{C}^{N \times 2}, ~ {\rm Tr}({\bf B} {\bf B}^H ) \leq 1}   &  ~ C_{\sf BF-ALAM}({\bf B},P),
\end{aligned}
\end{equation}
where
$$ C_{\sf BF-ALAM}({\bf B},P) = \min_{i= {1,\ldots,M}} \log ( 1 + P {\bf h}^H_i  {\bf B} {\bf B}^H {\bf h}_i ). $$
Note that $s(t)$ is assumed to be ideally channel-coded (just like in the beamforming case), and the constraint ${\rm Tr}({\bf B} {\bf B}^H ) \leq 1$ is equivalent to the total power constraint $\mathbb{E}[ \| {\bm X}(n) \|^2 ] /2 \leq P$.
In the next subsection, we will study how SDR can be employed to deal with the above achievable rate optimization problem.

\subsection{A Generalization of SDR for the Fixed Beamformed Alamouti Strategy}
\label{sec:fixed_BF_alam}
Our strategy for tackling \eqref{eq:mmf_alam} expands on the ideas used to reformulate the beamforming multicast achievable rate problem \eqref{eq:BF} into a rank-constrained SDP; see Section \ref{sec:formulation}.  To begin, observe that
$$ \mathbf{W}=\mathbf{B}\mathbf{B}^H \quad\Longleftrightarrow\quad \mathbf{W} \succeq \mathbf{0} \mbox{ and } \mbox{rank}(\mathbf{W}) \le 2. $$
Hence, Problem \eqref{eq:mmf_alam} can be equivalently formulated as
\begin{equation} \label{eq:mmf_alam_rank_sdr}
\begin{array}{rl}
   \displaystyle \max_{\mathbf{W} \in \mathbb{H}^N} &\displaystyle \min_{i= {1,\ldots,M}}  {\rm Tr}( \mathbf{W}\mathbf{h}_{i} \mathbf{h}_{i}^{H} )  \\
\text{s.t.}   &   {\rm Tr}( \mathbf{W} ) \leq 1, \, \mathbf{W} \succeq \mathbf{0}, \, \mbox{rank}(\mathbf{W}) \le 2.
\end{array}
\end{equation}
At this point, it is worth noting that the achievable rate problem for the beamforming scheme~\eqref{eq:mmf_rank_sdr} is a restriction of that for the beamformed Alamouti scheme~\eqref{eq:mmf_alam_rank_sdr}.  This suggests that our proposed design should have a performance no worse than that of the beamforming scheme.  In fact, as we shall see shortly, the worst-case performance gain can be quantified.

Now, upon removing the nonconvex rank constraint in \eqref{eq:mmf_alam_rank_sdr}, we obtain exactly the same convex relaxation as that of the fixed beamforming problem discussed in Section \ref{sec:fixed_BF}, namely, Problem (SDR).  Let $\mathbf{W}^\star$ denote an optimal solution to (SDR).  Since $\mathbf{W}^\star$ may not satisfy ${\rm rank}(\mathbf{W}^\star) \le 2$, we need to develop a procedure that can generate from $\mathbf{W}^\star$ a feasible solution to \eqref{eq:mmf_alam_rank_sdr}.  Moreover, since the generated solution need not be optimal for \eqref{eq:mmf_alam_rank_sdr} in general, we are interested in quantifying the approximation quality of such a solution.  To tackle these problems, we employ the SDR rank reduction theory (see, e.g., \cite{Jnl:Yongwei_rank,Jnl:SoYe2008}).  Let us begin with the following proposition:
\begin{Prop} \label{prop:exact-rk-red}
   Suppose that $M \le 8$.  Then, there is a polynomial-time procedure that can generate from $\mathbf{W}^\star$ an optimal solution $\hat{\mathbf{B}}$ to the fixed beamformed Alamouti problem \eqref{eq:mmf_alam}.
\end{Prop}
Proposition \ref{prop:exact-rk-red} can be established using~\cite[Claim 2]{MulticastSidiropoulos06} and \cite[Theorem 5.1]{Jnl:Yongwei_rank} (see also~\cite{Jnl:MagzineMaLUO} for an exposition of the latter).  It implies that the fixed beamformed Alamouti problem \eqref{eq:mmf_alam} can be optimally solved by SDR for instances with $8$ users or less.
By contrast, beamforming can guarantee the same result only for $3$ users or less; see Fact~\ref{fact:rank-1}(a).
Moreover, by the equivalence of (SDR) and (MC),
we arrive at the important conclusion that {\it fixed beamformed Alamouti space-time coding is a multicast capacity-optimal transmit strategy when there are no more than $8$ users.}

For the case where $M>8$, it may not be possible to generate an optimal solution to \eqref{eq:mmf_alam} from $\mathbf{W}^\star$ in polynomial time, as Problem \eqref{eq:mmf_alam} is NP-hard.  However, we can still generate a feasible solution to \eqref{eq:mmf_alam} using the following Gaussian randomization procedure:

\vspace{-0.4\baselineskip}

\begin{algorithm}[H]
\caption{Gaussian Randomization Procedure for \eqref{eq:mmf_alam}} \label{alg1}
\begin{algorithmic}[1]
\STATE Input: an optimal solution $\mathbf{W}^{\star}$ to (SDR), number of randomizations $L \ge 1$
\FOR {$j=1$ to $L$}
\STATE generate two independent random vectors $\bm{\xi}_1^j,\bm{\xi}_2^j \sim \mathcal{CN}(\mathbf{0}, \mathbf{W}^{\star})$ and define $\tilde{\mathbf{B}}_j = \frac{1}{\sqrt{2}} [ \begin{array}{cc} \bm{\xi}_1^j & \bm{\xi}_2^j \end{array} ]$;
\STATE let $\hat{\mathbf{B}}_j =  \tilde{\mathbf{B}}_j \Big/ \sqrt{ {\mbox{Tr}( \tilde{\mathbf{B}}_j\tilde{\mathbf{B}}_j^H )}}$
\ENDFOR
\STATE let $j^\star := \arg\max_{j=1,\ldots,L} \overline{\sf SNR}_{\rm min}(\hat{\mathbf{B}}_j\hat{\mathbf{B}}_j^H)$ (see \eqref{eq:worst-SNR} for the definition of $\overline{\sf SNR}_{\rm min}(\cdot)$)
\STATE Output: $\hat{\mathbf{B}} = \hat{\mathbf{B}}_{j^\star}$
\end{algorithmic}
\end{algorithm}

\vspace{-0.5\baselineskip}

Algorithm \ref{alg1} is a generalization of the Gaussian randomization procedure used for the SDR-based beamforming scheme~\cite{MulticastSidiropoulos06}.  Regarding its worst-case approximation performance, we have the following result, whose proof can be found in Appendix \ref{sec:sdr_alam_approx_pf}:
\begin{Theorem} \label{thm:sdr_alam_approx}
With probability at least $1-(5/6)^L$, the solution $\hat{\mathbf{B}}$ returned by Algorithm \ref{alg1} satisfies
$$ \overline{\sf SNR}_{\rm min}(\hat{\mathbf{B}}\hat{\mathbf{B}}^H) \ge \frac{\overline{\sf SNR}_{\rm min}({\bf W}^\star)}{12.22\sqrt{M}} = \frac{\rho_{\rm min}}{12.22\sqrt{M}}. $$
\end{Theorem}
Theorem \ref{thm:sdr_alam_approx} has two important implications.  First, with our fixed beamformed Alamouti scheme, the provable gap between the worst-user SNR and the best achievable worst-user SNR scales  only on the order of $\sqrt{M}$.  This is substantially better than the fixed beamforming case, where the provable gap scales on the order of $M$ (cf.~Fact~\ref{fact:rank-1}(b)).  Secondly, for $M>8$, the achievable rate gap of the SDR-based fixed beamformed Alamouti scheme relative to the multicast capacity is bounded above by
\ifconfver
\begin{align*}
& C_{\sf MC}(P) - C_{\sf BF-ALAM}( \hat{\bf B},P ) \\
&\leq \log \left( \frac{1+ \rho_{\rm min} P }{ 1 + \rho_{\rm min} P/(12.22\sqrt{M})} \right),
\end{align*}
\else
\begin{align*}
 C_{\sf MC}(P) - C_{\sf BF-ALAM}( \hat{\bf B},P ) \leq \log \left( \frac{1+ \rho_{\rm min} P }{ 1 + \rho_{\rm min} P/(12.22\sqrt{M})} \right),
\end{align*}
\fi
which for large $P$ is approximately equal to $\log(12.22\sqrt{M})$.  This is strictly better than that of the SDR-based fixed beamforming scheme for all $M>8$ (cf.~\eqref{eq:g_SDR_BF} in Fact~\ref{fact:rank-1}(b)).

Before we proceed, several remarks are in order.

\noindent{\it Remark 1:} The techniques we developed for proving Theorem \ref{thm:sdr_alam_approx} can be used to obtain approximation bounds for a fairly general class of rank constrained SDPs.  As such, they generalize the techniques in~\cite{MulticastLuo07}, which only apply to a certain class of rank-one constrained SDPs.

\noindent{\it Remark 2:} The approximation bound stated in Theorem \ref{thm:sdr_alam_approx} is only a worst-case bound.  In practice, the solution returned by Algorithm \ref{alg1} can have a much better performance.  This will be confirmed by our simulation results; see Section \ref{sec:sim}.

\noindent{\it Remark 3:} In view of the development of the fixed beamformed Alamouti scheme, it is natural to ask whether the techniques can be extended to deliver a ``rank-$n$'' beamforming scheme rather than just a ``rank-$2$'' scheme as in the Alamouti case.  Indeed, it is possible to extend the SDR techniques above to general $n$-dimensional orthogonal space-time bock codes (OSTBCs).
However, full rate OSTBCs do not exist for $n > 2$~\cite{Liang_OSTBC_03}, and the rate deduction (for $n > 2$) can significantly outweigh the gain obtained from ``rank-$n$'' beamforming.  For example, consider a fixed beamformed OSTBC for dimension $n=3$.  Since the maximal-rate OSTBC for $n=3$ is $3/4$~\cite{Liang_OSTBC_03},
the achievable rate should be formulated as
\[
C_{\sf BF-OSTBC}({\bf B},P) = \min_{i=1,\ldots,M} \frac{3}{4} \log\left( 1 +  P {\bf h}_i^H {\bf B}{\bf B}^H {\bf h}_i  \right),
\]
where ${\bf B} \in \mathbb{C}^{N \times 3}$, with ${\rm Tr}({\bf B}{\bf B}^H) \leq 4/3$.
Our SDR analysis can be extended to show that the solution $\hat{\mathbf{B}} \in \mathbb{C}^{N\times 3}$ generated by a certain Gaussian randomization procedure will satisfy $\overline{\sf SNR}_{\rm min}( \hat{\bf B}\hat{\bf B}^H ) \geq \overline{\sf SNR}_{\rm min}( {\bf W}^\star ) / \mathcal{O}( M^{1/3} )$ with high probability, which further improves upon the result in the beamformed Alamouti case (cf.~Theorem~\ref{thm:sdr_alam_approx}).  However, this effective SNR gain can easily be compromised by the $3/4$ factor in the overall achievable rate, especially for large $P$.  The issue of having no full rate OSTBCs for $n > 2$ makes the further development of beamformed OSTBCs unattractive.

\section{Combining the SBF and Alamouti Strategies}
\label{sec:SBF_alam}

In this section we present our last technical contribution, namely, to demonstrate how the two physical-layer multicasting strategies proposed in the previous sections can be combined to yield SBF Alamouti schemes, and to analyze the performance of the resulting schemes.

\subsection{Main Results}

The system model of the SBF Alamouti strategy is identical to that of the fixed beamformed Almaouti strategy in Section~\ref{sec:alam_model},
except that the transmit space-time code blocks in \eqref{eq:X_codeblk} are changed to
\[
{\bm X}(n) = \sqrt{P} {\bf B}(n) \mathbf{C}( {\bm s}(n) ),
\]
where ${\bf B}(n) \in \mathbb{C}^{N \times 2}$ is a random-in-block beamforming matrix.
In other words, we take the Alamouti space-time structure
while randomizing the beamforming matrix, just as in SBF.
Following the same derivations as in Section~\ref{sec:alam_model} and adopting the SBF formulation in Section~\ref{sec:SBF},
we can express the multicast achievable rate of an SBF Alamouti scheme as
\begin{equation} \label{eq:C_SBF_alam}
C_{\sf SBF-ALAM}(P)=  \min_{i=1,\ldots,M}\mathbb{E}_{\bm{B} \sim \mathcal{D}}[\log (1+ P {\bf h}_{i}^{H}{ \bm{B}\bm{B}^{H} } {\bf h}_{i} )],
\end{equation}
and the corresponding achievable rate gap as
$$ g_{\sf SBF-ALAM}(P) =  C_{\sf MC}(P) - C_{\sf SBF-ALAM}(P). $$
Here, $\bm{B} \in \mathbb{C}^{N \times 2}$ is a random matrix, and $\mathcal{D}$ denotes its corresponding beamformer matrix distribution, which must satisfy $\mathbb{E}_{\bm{B} \sim \mathcal{D}}[{\rm Tr}(\bm{B} \bm{B}^H )] \leq 1$.
The SBF Alamouti schemes to be proposed follow the same spirit as the original SBF schemes.
To describe them, let $\bm{B} = [~ \bm{w}_1, \bm{w}_2 ~]$, and denote
\[
\bar{\bm{w}} = \begin{bmatrix} \bm{w}_1 \\ \bm{w}_2 \end{bmatrix}, \quad
\bar{\bf L} = \frac{1}{\sqrt{2}} \begin{bmatrix} {\bf L} & {\bf 0} \\ {\bf 0} & {\bf L} \end{bmatrix}, \quad
\bar{\bm{\alpha}} = \begin{bmatrix} \bm{\alpha}_1 \\ \bm{\alpha}_2 \end{bmatrix},
\]
where $\bm{\alpha}_1, \bm{\alpha}_2 \sim \mathcal{CN}({\bf 0},{\bf I}_r)$ are independent random vectors,
$r= {\rm rank}({\bf W}^\star)$,
and ${\bf L} \in \mathbb{C}^{r \times N}$ is a square root decomposition of ${\bf W}^\star$ satisfying ${\bf L}^H {\bf L} = {\bf W}^\star$.  We propose the following three schemes:
\begin{itemize}
\item {\it Gaussian SBF Alamouti scheme:} \ $\bar{\bm{w}} = \bar{\bf L}^H \bar{\bm{\alpha}}$;
\item {\it Elliptic SBF Alamouti scheme:} \ $\bar{\bm{w}} = \bar{\bf L}^H \bar{\bm{\alpha}} / (\| \bar{\bm{\alpha}} \| / \sqrt{2r}  )$;
\item {\it Bingham SBF Alamouti scheme:} \ $\bar{\bm{w}} = \bar{\bf L}^H \bar{\bm{\alpha}} / \| \bar{\bf L}^H \bar{\bm{\alpha}} \|$.
\end{itemize}
The Gaussian SBF Alamouti scheme satisfies the multicast capacity-optimal transmit covariance property
$\mathbb{E}[ \bm{B} \bm{B}^H ]= \mathbb{E}[ \bm{w}_1 \bm{w}_1^H ] + \mathbb{E}[ \bm{w}_2 \bm{w}_2^H ] = {\bf W}^\star$, as one can easily verify.  The elliptic SBF Alamouti scheme also satisfies this property, as implied by Fact~\ref{fact:ellip}.  On the other hand, the Bingham SBF Alamouti scheme may not satisfy the transmit covariance property.  The following theorem summarizes our main results:
\begin{Theorem} \label{thm:SBF_alam}
The achievable rate gaps of the Gaussian, elliptic and Bingham SBF Alamouti schemes satisfy
\begin{align*}
g_{\sf SBF-ALAM}^{\sf Gauss}(P) & \leq \log(2) + \gamma - 1 = 0.2703, \\
g_{\sf SBF-ALAM}^{\sf Ellip}(P) & \leq \sum_{k=1}^{2r-1} \frac{1}{k} - \log(r) - 1,  \\
g_{\sf SBF-ALAM}^{\sf Bing}(P) & \leq \sum_{k=1}^{2r-1} \frac{1}{k} - \log(r) - 1
\end{align*}
for all $P \geq 0$, respectively.  For the Gaussian and elliptic cases, the bounds are tight when $P \rightarrow \infty$.
\end{Theorem}
The proof of Theorem~\ref{thm:SBF_alam} will be provided in the next subsection.  Similar to the analysis of the SBF schemes, it can be shown that $\sum_{k=1}^{2r-1} \frac{1}{k} - \log(r) - 1$ increases with $r$, and that $\sum_{k=1}^{2r-1} \frac{1}{k} - \log(r) - 1$ approaches $\log(2) + \gamma - 1$ as $r \rightarrow \infty$.  This means that the worst-case rate gaps of the elliptic and Bingham SBF Alamouti schemes are no worse than that of the Gaussian SBF Alamouti scheme, and can be much better for smaller $r$.
Table~\ref{tab:2} shows the rate gap values of the elliptic and Bingham SBF Alamouti schemes for various $r$.
Theorem~\ref{thm:SBF_alam} also provides the vital implication
that the three SBF Alamouti schemes narrow the worst-case rate loss down to $0.39$~bits/s/Hz ($0.2703/\log(2)= 0.39$),
again, irrespective of any factors.


\begin{table}[h]
\caption{\protect{The worst-case rate gap of the elliptic and Bingham SBF Alamouti schemes}}
\vspace*{-.5\baselineskip}
\renewcommand{\arraystretch}{1.3} \centering
\begin{tabular}{|c|c|c|c|c|c|}
\hline $r$ &~~~$1$~~~~& $2$& $3$& $\ldots$ & $\infty$
\\\hline rate gap in nats & $0$ & $0.1402$ &$0.1847$& $\ldots$&$0.2703$
\\\hline rate gap in bits & $0$ & $0.2023$ &$0.2665$& $\ldots$&$0.39$\\\hline
\end{tabular}
\label{tab:2}
\vspace*{-\baselineskip}
\end{table}

In preparation for the proof of Theorem \ref{thm:SBF_alam}, let us observe that the SBF Alamouti multicast achievable rate $C_{\sf SBF-ALAM}(P)$ in \eqref{eq:C_SBF_alam} can be expressed as
$$ 
C_{\sf SBF-ALAM}(P)=  \min_{i=1,\ldots,M}\mathbb{E}_{\xi_i}[\log (1+ \xi_i \rho_i P ) ],
$$ 
where $\rho_i$ is defined in~\eqref{eq:rho}, and
\begin{equation} \label{eq:xi_i_2}
\xi_i = \frac{ | {\bf h}_{i}^{H} \bm{w}_1 |^2 + | {\bf h}_{i}^{H} \bm{w}_2 |^2 }{\rho_i }, \quad i=1,\ldots,M.
\end{equation}
In particular, the distributions of the random variables $\xi_1,\ldots,\xi_M$ will play an important role in our analysis.

\subsection{Proof of Theorem~\ref{thm:SBF_alam}: The Gaussian Case}
For the Gaussian SBF Alamouti scheme, it is routine to show that the $\xi_i$'s in \eqref{eq:xi_i_2} follow a chi-square distribution with unit mean and $4$ degrees of freedom.  Thus, by Fact~\ref{fact:SBF_basic}(a), we have
$$ 
C_{\sf SBF-ALAM}^{\sf Gauss}(P)=  \mathbb{E}_{\xi }[\log (1 + \xi\rho_{\rm min} P )],
$$ 
where $\xi \sim \xi_i$ for any $i$.  Moreover, by Fact~\ref{fact:SBF_basic}(b), the achievable rate gap $g_{\sf SBF-ALAM}^{\sf Gauss}(P)= C_{\sf MC}(P) - C_{\sf SBF-ALAM}^{\sf Gauss}(P)$ is nondecreasing in $P \geq 0$.  The claim for the Gaussian SBF Alamouti rate gap in Theorem~\ref{thm:SBF_alam} now follows from the following proposition, whose proof can be found in Section 3.1 of the companion technical report~\cite{CompTechRep}:
\begin{Prop}
For any $P>0$,
$$ C_{\sf SBF-ALAM}^{\sf Gauss}(P)=  \left( 1 - \frac{2}{\rho_{\rm min} P} \right) e^{\tfrac{2}{\rho_{\rm min} P}} E_1 \left(\frac{2}{\rho_{\rm min} P}\right) + 1. $$
Consequently, we have
$$ \lim_{P \rightarrow \infty} g_{\sf SBF-ALAM}^{\sf Gauss}(P) = \log(2) + \gamma - 1. $$
\end{Prop}


\subsection{Proof of Theorem~\ref{thm:SBF_alam}: The Elliptic Case}

For the elliptic SBF Alamouti scheme, we compute
$$ \bar{\bm{w}} = \frac{\sqrt{r}}{\| \bar{\bm{\alpha}} \|} \begin{bmatrix} {\bf L}^H\bm{\alpha}_1 \\ {\bf L}^H\bm{\alpha}_2 \end{bmatrix}. $$
Together with \eqref{eq:xi_i_2}, this gives
\begin{equation} \label{eq:xi_i_ellip}
\xi_i = \frac{| (\sqrt{r/\rho_i} {\bf L}{\bf h}_{i})^H \bm{\alpha}_1 |^2 + | (\sqrt{r/\rho_i} {\bf L}{\bf h}_{i})^H \bm{\alpha}_2 |^2}{\| \bm{\alpha}_1 \|^2 + \| \bm{\alpha}_2 \|^2}.
\end{equation}
In particular, if we take ${\bf u} = \sqrt{r/\rho_i} {\bf L}{\bf h}_i$ and $l=2$ in Lemma~\ref{lem:ellip_dist_gen}, differentiate the corresponding CDF w.r.t.~$t$ and observe that $\|{\bf u}\|^2=r$, we obtain the following result:
\begin{Prop} \label{prop:ellip_alam_dist_fn}
  Consider the elliptic SBF Alamouti scheme.  The PDF of $\xi_i$, where $i=1,\ldots,M$, is given by
  $$ 
     p_{\xi_i}(t) = \frac{(2r-1)(2r-2)}{r} \cdot \frac{t}{r} \left( 1-\frac{t}{r} \right)^{2r-3} \,\,\,\mbox{for } 0 \le t \le r,
  $$ 
where $r={\rm rank}( {\bf W}^\star)$.
\end{Prop}
Proposition \ref{prop:ellip_alam_dist_fn} implies that the $\xi_i$'s in \eqref{eq:xi_i_ellip} are identically distributed.  Hence, by Fact~\ref{fact:SBF_basic}(a), we have $C_{\sf SBF-ALAM}^{\sf Ellip}(P) = \mathbb{E}_{\xi }[\log ( 1 + \xi\rho_{\rm min} P )]$, where $\xi \sim \xi_i$ for any $i$.  Moreover, since $\mathbb{E}[ \xi_i ]=1$ for all $i$, by Fact~\ref{fact:SBF_basic}(b), the achievable rate gap $g_{\sf SBF-ALAM}^{\sf Ellip}(P)= C_{\sf MC}(P) - C_{\sf SBF-ALAM}^{\sf Ellip}(P)$ is nondecreasing in $P\ge0$.  The claim for the elliptic SBF Alamouti rate gap in Theorem~\ref{thm:SBF_alam} now follows from the following proposition, whose proof can be found in Section 3.2 of the companion technical report~\cite{CompTechRep}:
\begin{Prop}
For any $P>0$,
\ifconfver
\begin{align*}
   & C_{\sf SBF-ALAM}^{\sf Ellip}(P) \\ 
   &= (2r-1) \left( 1+\frac{1}{r\rho_{\rm min}P} \right)^{2r-2} \Bigg[ \log(1+r\rho_{\rm min}P) \\
   & \left. \hspace{0.5in}  - \sum_{k=1}^{2r-2} \frac{1}{k} - \sum_{k=1}^{2r-2} {2r-2 \choose k} \frac{(-1)^k}{k(1+r\rho_{\rm min}P)^k} \right] \\
   &\quad- (2r-2)\left( 1+\frac{1}{r\rho_{\rm min}P} \right)^{2r-1} \Bigg[ \log(1+r\rho_{\rm min}P) \\
   & \left. \hspace{0.5in} - \sum_{k=1}^{2r-1} \frac{1}{k} - \sum_{k=1}^{2r-1} {2r-1 \choose k} \frac{(-1)^k}{k(1+r\rho_{\rm min}P)^k} \right]. 
\end{align*}
\else
\begin{align*}
   & C_{\sf SBF-ALAM}^{\sf Ellip}(P) \\
   & = (2r-1) \left( 1+\frac{1}{r\rho_{\rm min}P} \right)^{2r-2} \left[ \log(1+r\rho_{\rm min}P)  - \sum_{k=1}^{2r-2} \frac{1}{k} - \sum_{k=1}^{2r-2} {2r-2 \choose k} \frac{(-1)^k}{k(1+r\rho_{\rm min}P)^k} \right] \\
   & \quad - (2r-2)\left( 1+\frac{1}{r\rho_{\rm min}P} \right)^{2r-1} \left[ \log(1+r\rho_{\rm min}P)  - \sum_{k=1}^{2r-1} \frac{1}{k} - \sum_{k=1}^{2r-1} {2r-1 \choose k} \frac{(-1)^k}{k(1+r\rho_{\rm min}P)^k} \right]. 
\end{align*}
\fi
Consequently, we have
$$ \lim_{P \rightarrow \infty} g_{\sf SBF-ALAM}^{\sf Ellip}(P) = \sum_{k=1}^{2r-1} \frac{1}{k} - \log(r) - 1. $$
\end{Prop}

\subsection{Proof of Theorem~\ref{thm:SBF_alam}: The Bingham Case}

By extending the proof of Proposition~\ref{prop:bing_rate}, we show that the Bingham SBF Alamouti rate of user $i$ is given by
\begin{align}
C_{{\sf SBF-ALAM},i}^{\sf Bing}(P) &= \mathbb{E}_{\bm{B}}[ \log( 1 + P {\bf h}_{i}^{H}{ \bm{B}\bm{B}^{H} } {\bf h}_{i} ) ] \nonumber \\
&= \log( 1 + \rho_i P ) +
    \bar{\varphi}\left( \frac{\bm{\mu}_i}{ \bm{\mu}_i^T {\bf 1} } \right)
    - \bar{\varphi}\left( \bm{\lambda} \right), \label{eq:bing_rate_2}
\end{align}
where $\bm{\lambda}$ contains the $r$ positive eigenvalues of ${\bf W}^\star$,  $\bm{\mu}_i$ contains the eigenvalues of the matrix ${\bf A}_i = {\bf L} ( {\bf I}_N + P {\bf h}_i {\bf h}_i^H ){\bf L}^H$,
and
$\bar{\varphi}: \mathbb{R}^r \rightarrow \mathbb{R}$ is defined by
\begin{equation} \label{eq:bar_varphi}
\bar{\varphi}\left( {\bf d} \right) = \mathbb{E}_{\bm{\zeta}_1,\bm{\zeta}_2}\left[ \log\left( \sum_{k=1}^r d_k \frac{\zeta_{1,k} +  \zeta_{2,k}}{2} \right)   \right].
\end{equation}
Here, $\bm{\zeta}_1$ and $\bm{\zeta}_2$ are independent random vectors with i.i.d. unit-mean exponentially distributed components.
Note that the difference between the above results and Proposition~\ref{prop:bing_rate} lies in \eqref{eq:bar_varphi}.
Although it is possible to derive an explicit expression for \eqref{eq:bar_varphi} by applying the result in Proposition~\ref{fact:sumofexp}, such an expression will be too complicated for analysis purposes.
Thus, we turn to stochastic majorization techniques to analyze the function $\bar{\varphi}$.  Using Fact~\ref{fact:varphi}, we deduce that
\begin{align}
\bar{\varphi}\left( {\bf e}_1 \right) & \leq \bar{\varphi}\left( {\bf d} \right) \leq \bar{\varphi}\left( \tfrac{1}{n} {\bf 1} \right), \nonumber \\
\bar{\varphi}\left( \tfrac{1}{n} {\bf 1} \right) & = \sum_{k=1}^{2n-1} \frac{1}{k} - \log(2n) - \gamma   \label{eq:varphi_eq}
\end{align}
for any ${\bf d}= (d_1,\ldots,d_n) \ge {\bf 0}$ with $\sum_{k=1}^n d_k = 1$.
Note that \eqref{eq:varphi_eq} is obtained from the relation $\bar{\varphi}( {\bf 1}/n ) = \varphi( [ {\bf 1}^T \, {\bf 1}^T ]^T /2n )$. Applying the above inequalities to \eqref{eq:bing_rate_2} yields
\ifconfver
\begin{align*}
& \mathbb{E}_{\bm{B}}[ \log( 1 + P {\bf h}_{i}^{H}{ \bm{B}\bm{B}^{H} } {\bf h}_{i} ) ] \\
& \geq \log( 1 + \rho_i P ) + \bar{\varphi}( {\bf e}_1 ) - \bar{\varphi}\left(  \tfrac{1}{r} {\bf 1} \right) \\ 
& = \log( 1 + \rho_i P ) + (1 - \log(2) - \gamma) \\
&\quad - \left( \sum_{k=1}^{2r-1} \frac{1}{k} - \log(2r) - \gamma \right) \\ 
& \ge \log( 1 + \rho_{\rm min} P ) + \log(r) +1 - \sum_{k=1}^{2r-1} \frac{1}{k}.
\end{align*}
\else
\begin{align*}
\mathbb{E}_{\bm{B}}[ \log( 1 + P {\bf h}_{i}^{H}{ \bm{B}\bm{B}^{H} } {\bf h}_{i} ) ]
& \geq \log( 1 + \rho_i P ) + \bar{\varphi}( {\bf e}_1 ) - \bar{\varphi}\left(  \tfrac{1}{r} {\bf 1} \right) \\ 
& = \log( 1 + \rho_i P ) + (1 - \log(2) - \gamma)
 - \left( \sum_{k=1}^{2r-1} \frac{1}{k} - \log(2r) - \gamma \right) \\ 
& \ge \log( 1 + \rho_{\rm min} P ) + \log(r) +1 - \sum_{k=1}^{2r-1} \frac{1}{k}.
\end{align*}
\fi
Since $C_{\sf MC}(P) = \log( 1+\rho_{\rm min}P )$, we conclude that the achievable rate gap $g_{\sf SBF-ALAM}^{\sf Bing}(P)= C_{\sf MC}(P) - C_{\sf SBF-ALAM}^{\sf Bing}(P)$ satisfies
$$ g_{\sf SBF-ALAM}^{\sf Bing}(P) \le \sum_{k=1}^{2r-1} \frac{1}{k} - \log(r) - 1, $$
as desired.

\section{Simulation Results}
\label{sec:sim}

This section presents simulation results for the proposed multicast SBF schemes.
Unless specified, all the results to be shown were obtained from $1,000$ trials of randomly generated channel realizations,
where ${\bf h}_i \sim \mathcal{CN}({\bf 0},{\bf I}_N)$ for each trial.
The SDR-based beamforming scheme, which will be benchmarked, is implemented by the Gaussian randomization procedure
(see \cite[Table II, with ``\textsf{randC}'' generation]{MulticastSidiropoulos06})
with $30MN$ number of randomizations.
For convenience, we shall refer to the SDR-based beamforming scheme
(resp. SDR-based beamformed Alamouti scheme)
as ``beamforming'' (resp. ``beamformed Alamouti'').
To illustrate how good a scheme can utilize CSIT, we will also evaluate the multicast achievable rate of the open-loop strategy, which is the multicast rate in \eqref{eq:MC} when the transmit covariance is fixed as ${\bf W}= \frac{1}{N}{\bf I}$~\cite{Conferece:CapacityLimitsLuo06,Jnl:Subsection_Love_08}.

\subsection{Multicast Achievable Rate Performance}

\begin{figure}[htp]
\begin{center}
\subfigure[][]{\resizebox{0.485\textwidth}{!}{\includegraphics{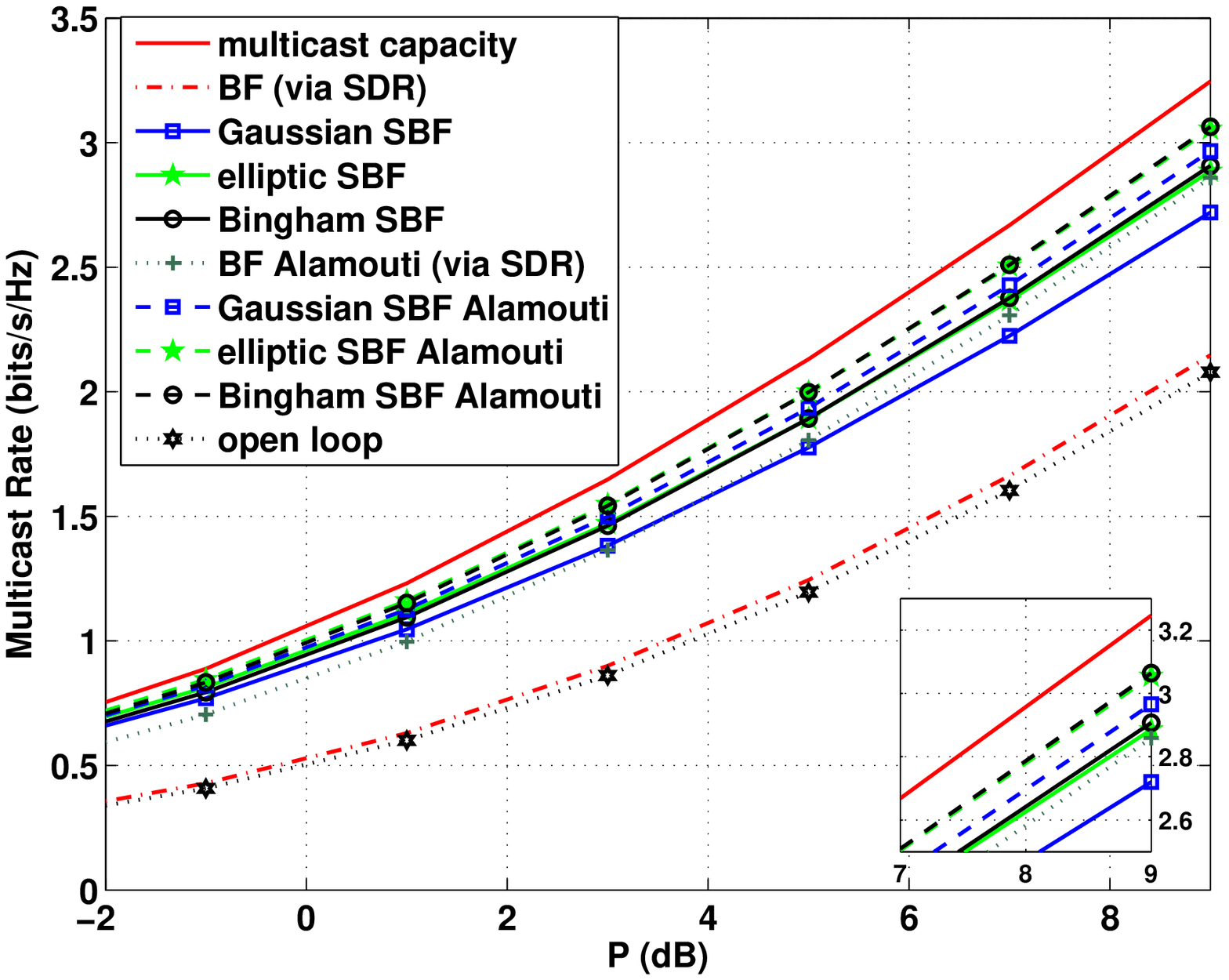}}}
\subfigure[][]{\resizebox{0.485\textwidth}{!}{\includegraphics{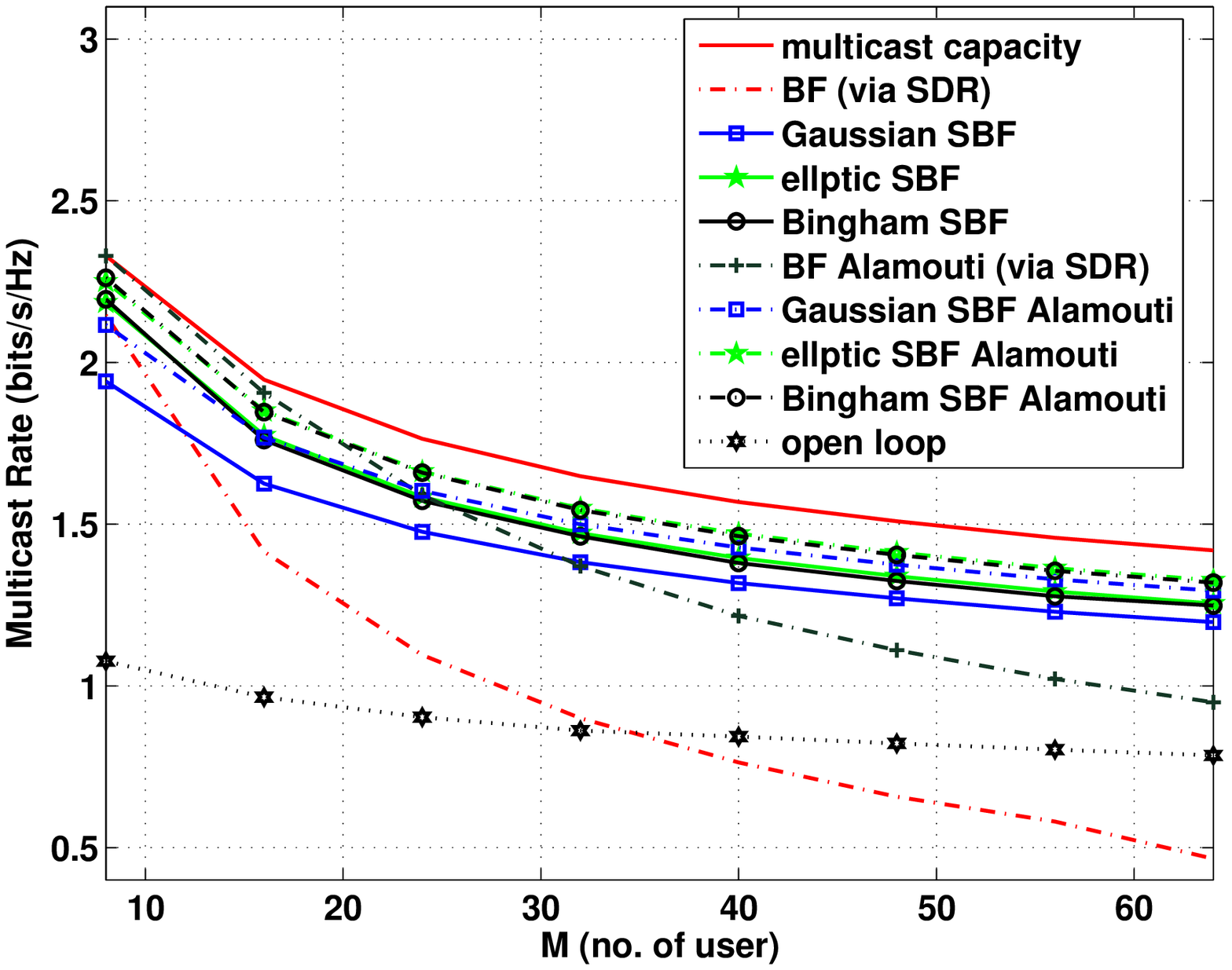}}}
\end{center}
\vspace*{-\baselineskip}
\caption{Multicast achievable rates of the various multicast schemes.}
\label{fig:SBF_rate}
\end{figure}

Fig.~\ref{fig:SBF_rate}(a) plots the multicast achievable rates of the various schemes w.r.t. the power $P$,
when the number of transmit antennas and users are $N= 8$ and $M= 32$, resp.
Note that the rates shown are averages of all the trials.
One can see that the SBF schemes substantially outperform beamforming.
In fact, beamforming shows very little rate advantage over the open-loop strategy in this many-user setting.
However, this is not the case with SBF.
At this point, it should be added that in all the trials run,
we found $64$ times of having ${\rm rank}({\bf W}^\star)= 2$,
$846$ times of ${\rm rank}({\bf W}^\star)= 3$,
and $90$ times of ${\rm rank}({\bf W}^\star)= 4$.
Based on our empirical observation, the performance difference between SBF and beamforming is attributed to the higher rank instances.
By examining Fig.~\ref{fig:SBF_rate}(a) carefully,
we see that the SBF rate gaps relative to the multicast capacity are no greater than $0.5$~bits/s/Hz (under the tested range $-2{\rm dB} \leq P \leq 9{\rm dB}$), which fall well within the $0.8314$~bits/s/Hz worst-case bound proven in Theorems~\ref{thm:SBF_Gaus}-\ref{thm:SBF_Bing}.
The elliptic and Bingham SBF schemes yield very similar rate performance, and they perform better than the Gaussian SBF.
For the beamformed Alamouti scheme,
its rate is lower than the SBF schemes for $P \leq 3$dB, but catches up as $P$ increases.
For the SBF Alamouti schemes, they exhibit similar rate performance behavior compared to their SBF counterparts, but with improved rate values.
In particular, upon a closer inspection of Fig.~\ref{fig:SBF_rate}(a), we see that the SBF Alamouti rate gaps are no greater than $0.25$~bits/s/Hz, which is well within the $0.39$~bits/s/Hz worst-case bound claimed in Theorem~\ref{thm:SBF_alam}.

Fig.~\ref{fig:SBF_rate}(b) plots the multicast rates w.r.t. the number of users $M$,
when $N= 8$ and $P= 3$dB.
Beamforming is seen to provide good performance for small $M$, say, $M \leq 11$;
numerically it is noted that SDR has a higher chance to give rank-one solutions for small $M$.
However, we also see that the rate gap of beamforming (relative to the multicast capacity) widens as $M$ increases.
In particular, beamforming has no advantage over the open-loop strategy for $M > 32$.
In comparison, the SBF rate gaps, with and without Alamouti, are quite constant w.r.t. $M$,
which agrees well with the constant rate gap result in Theorems~\ref{thm:SBF_Gaus}-\ref{thm:SBF_Bing} and Theorem~\ref{thm:SBF_alam}.
They are also better than the open-loop multicast rate even for $M= 64$.
This demonstrates the superiority of the SBF strategy when there is a large number of users.
Like beamforming, the beamformed Alamouti scheme exhibits a rate gap widening effect as $M$ increases.  Nevertheless, the beamformed Alamouti rate is much better than that of beamforming---in fact, the former is seen to be better than all the SBF-based schemes for $M \leq 19$.

\subsection{Coded BER Performance}

\begin{figure}[htp]
\centering
\begin{center}
\subfigure[][$M=32$]{\resizebox{0.485\textwidth}{!}{\includegraphics{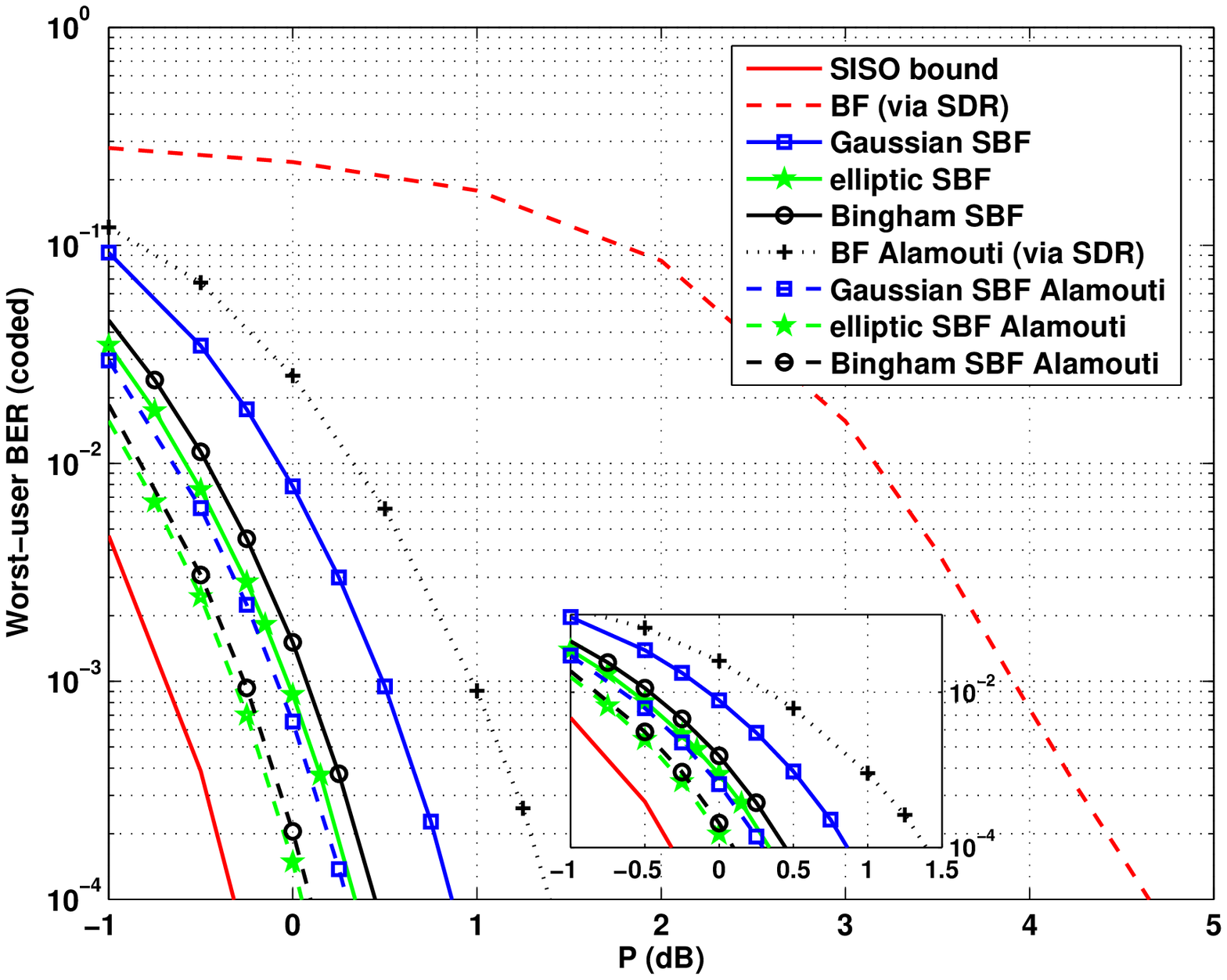}}} \subfigure[][$M=16$]{\resizebox{0.485\textwidth}{!}{\includegraphics{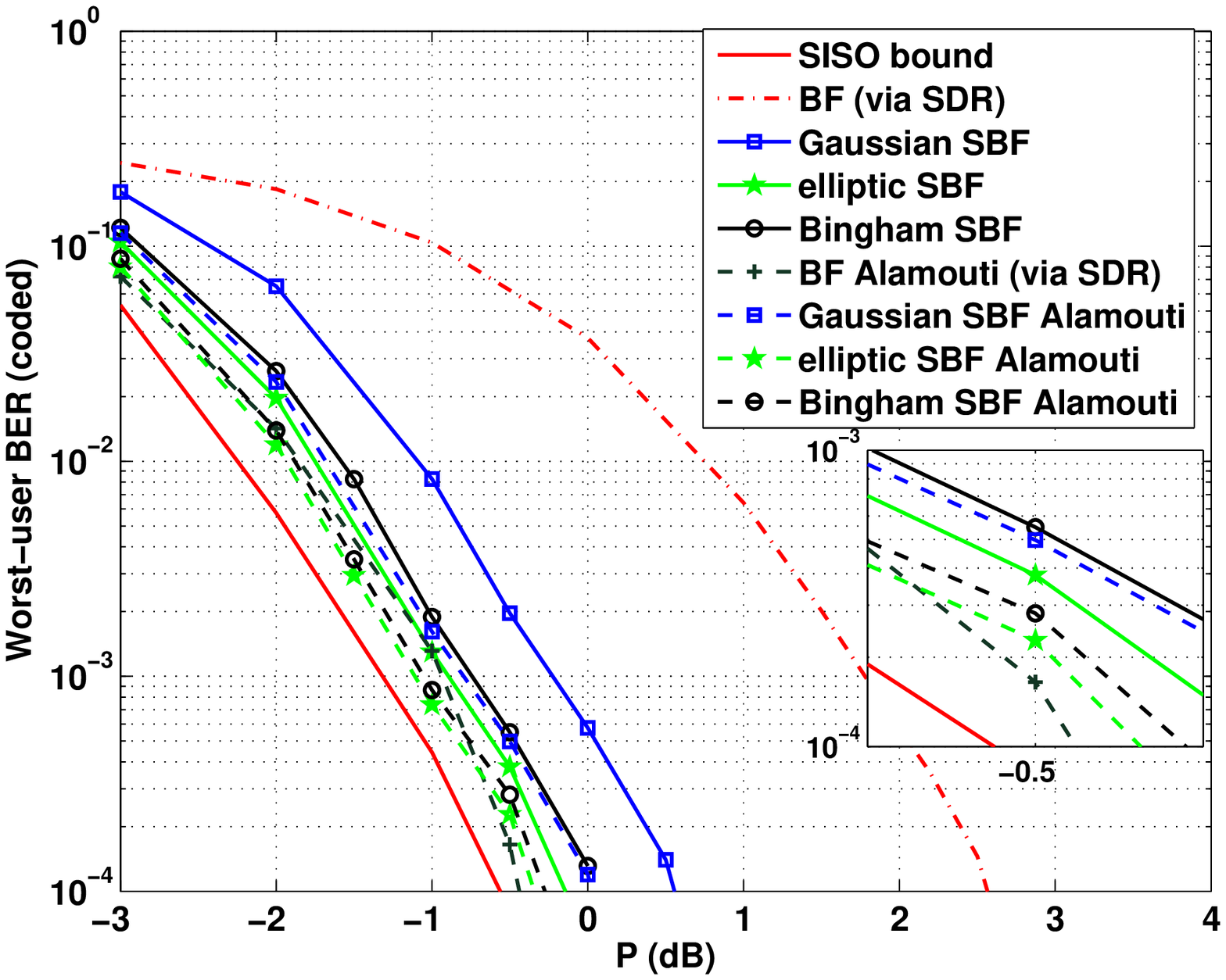}}}
\end{center}
\vspace*{-.5\baselineskip}
\caption{The worst-user BER performance of the various multicast schemes. QPSK; rate-$1/3$ Turbo code; $T=1440$.}
\label{fig:BER}
\end{figure}

Next, we physically realize the various schemes and evaluate their bit error rates (BERs).
The simulation setting is the same as that in Fig.~\ref{fig:SBF_rate}(a).
All the schemes adopt a rate-$1/3$ Turbo code with an information length of $960$ bits---which is used in IEEE $802.16$e~\cite{Std:16e}---as the channel coding scheme (with $10$ decoding iterations).
The modulation is Gray-coded QPSK.
There are totally $1440$ symbols in one frame, i.e., $T=1440$.
We ran $1,000$ independent data frames for each SNR point, so that the BER reliability level is $10$e$-5$.
We evaluated the worst-user BERs,
and the results are shown in Fig.~\ref{fig:BER}(a).
Note that in the figure,
``SISO bound'' is not a real multicast simulation.  It was obtained by running a single-user SISO system with SNR $\rho_{\rm min} P$ and with the same channel coding scheme.
It is expected that even a multicast capacity-achieving scheme, if it exists, should perform no better than the SISO bound.  Thus, the latter serves as a good BER baseline index.
Fig.~\ref{fig:BER}(a) demonstrates that the proposed schemes are much better than beamforming, this time in BER.
For example, fixing BER$= 10$e$-5$,
the elliptic SBF Alamouti scheme achieves an SNR gain of more than $4.5$dB relative to beamforming,
and is less than $0.5$dB away from the SISO bound.
Also, the BER performance ranking of the various schemes appears to be quite consistent with their achievable rate counterpart in Fig.~\ref{fig:SBF_rate}(a).
In Fig.~\ref{fig:BER}(b) we show another result where the number of users $M$ is reduced to $16$.
Beamforming is seen to provide better BER performance in comparison to the case of $M=32$, although SBF still performs better than beamforming.
Moreover, the beamformed Alamouti scheme now shows much improved performance.
This demonstrates that the beamformed Alamouti scheme can have competitive performance for smaller number of users.

\begin{figure}[htp]
\centering
\begin{center}
\resizebox{0.485\textwidth}{!}{\includegraphics{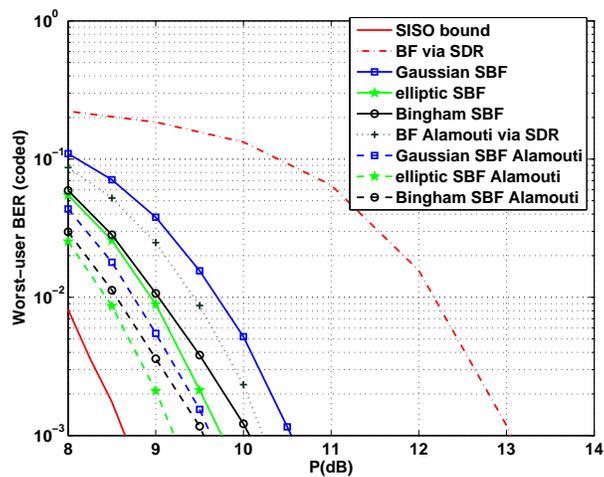}} 
\end{center}
\vspace*{-.5\baselineskip}
\caption{The worst-user BER performance of the various multicast schemes. $M= 32$; $16$-QAM; rate-$1/2$ Turbo code; $T=144$.}
\label{fig:BER2}
\end{figure}

In the previous simulation, we employ a relatively long frame length, namely,
$T=1440$,
which may be too long to some wireless scenarios.
For example, in the LTE standard, the frame length may be as small as $168$ symbols~\cite{3GPP_TS_36_211_9_10_0_0}.
In this simulation, a shorter frame length $T$ is considered.
We employ similar simulation settings as above,
except that we now use $16$-QAM and a rate-$1/2$ Turbo code with an information length of $288$ bits.
The consequent frame length is $T= 144$.
Also, $100$ independent data frames for each SNR point were run.
The results, shown in Fig.~\ref{fig:BER2}, illustrate that the performance of the various proposed schemes are generally consistent compared to the previous large frame-length BER simulations.

Before we close this section, we would like to draw the reader's attention to the companion technical report~\cite{CompTechRep}, which contains more simulation comparisons.

\section{Conclusion}
\label{sec:conclusion}

In this paper we established several new physical-layer multicasting schemes using stochastic beamforming and beamformed Alamouti space-time coding.
The proposed schemes are efficient to implement---the receiver sides require only symbol-by-symbol receiver processing, followed by a standard channel decoding operation.
We characterized the performance of the proposed schemes by means of theoretical analysis,
and showed that the proposed schemes have provably better multicast achievable rate scaling than the existing SDR-based multicast beamforming scheme w.r.t. the number of users.
We also demonstrated by simulations that the proposed schemes can outperform SDR-based beamforming quite significantly in terms of BERs under channel-coded, many-user settings.
%
As a future direction, it would be interesting to extend the present results to other scenarios.

\appendix

\section{Appendix}

\subsection{Proof of Fact~\ref{fact:SBF_basic}(b)}
\label{sec:A}

By Fact~\ref{fact:SBF_basic}(a), we have
$g_{\sf SBF}(P) = \log( 1 + \rho_{\rm min} P ) - \mathbb{E}_{\xi}[ \log( 1 + \xi \rho_{\rm min} P ) ]$.
Differentiating
$g_{\sf SBF}(P)$
w.r.t. $P$ yields
\begin{equation}
g^\prime_{\sf SBF}(P) = \left( \frac{1}{ 1 + \rho_{\rm min} P } - \mathbb{E}_{\xi} \left[ \frac{ \xi}{1+ \rho_{\rm min} P \xi }  \right] \right) \rho_{\rm min}.
\label{eq:gp_P}
\end{equation}
One can easily verify that for a fixed $P\ge0$, the function $\xi \mapsto \xi/( 1 + \rho_{\rm min} P \xi )$ is concave in $\xi \geq 0$.  Upon applying Jensen's inequality to \eqref{eq:gp_P} and using the fact that $\mathbb{E}_\xi[ \xi ] =1$, we get
\begin{equation*}
g^\prime_{\sf SBF}(P) \geq \left( \frac{1}{ 1 + \rho_{\rm min} P } -  \frac{ \mathbb{E}_{\xi} [ \xi ] }{1+ \rho_{\rm min} P \mathbb{E}_{\xi} [ \xi ] } \right) \rho_{\rm min} = 0,
\label{eq:gp_P_2}
\end{equation*}
i.e., $g_{\sf SBF}(P)$ is nondecreasing in $P \geq 0$.

\subsection{Proof of Lemma \ref{lem:ellip_dist_gen}}
\label{sec:B}

Since the distribution of $\bm{\alpha}_i$ is rotationally invariant (i.e., $\bm{\alpha}_i$ and ${\bf U}\bm{\alpha}_i$ have the same distribution for any fixed unitary matrix ${\bf U}$), we may assume without loss that ${\bf u} = (\|{\bf u}\|,0,\ldots,0)$.  Then, for any $t \ge 0$, we have
\begin{align*}
& \Pr( \eta({\bf u}) \le t ) = \Pr\left( \sum_{i=1}^l \| {\bf u} \|^2 | \alpha_{i1}|^2 \le t \sum_{i=1}^l \| \bm{\alpha}_i \|^2 \right) \\
&=  \Pr\left( \left( \|{\bf u}\|^2 - t \right) \sum_{i=1}^l |\alpha_{i1}|^2 \le t \sum_{i=1}^l\sum_{j=2}^r |\alpha_{ij}|^2 \right).
\end{align*}
By definition, $\chi_{2l}^2 = 2 \sum_{i=1}^l |\alpha_{i1}|^2$ and $\tilde{\chi}_{2l(r-1)}^2 = 2 \sum_{i=1}^l\sum_{j=2}^r |\alpha_{ij}|^2$ are independent chi-square random variables with $2l$ and $2l(r-1)$ degrees of freedom, resp.  It follows that
\begin{align}
& \Pr( \eta({\bf u}) \le t ) = \Pr\left( \frac{\tilde{\chi}_{2l(r-1)}^2 }{\chi_{2l}^2} \ge \frac{ \|{\bf u}\|^2 -t }{t} \right) \nonumber \\
&=  \Pr\left( \frac{\tilde{\chi}_{2l(r-1)}^2 / 2l(r-1)}{\chi_{2l}^2/2l} \ge \frac{ \|{\bf u}\|^2 -t }{t(r-1)} \right).\label{eq:gen_ellip_cdf}
\end{align}
Now, the non-negative random variable
$$ F_{2l(r-1),2l} = \frac{\tilde{\chi}_{2l(r-1)}^2 / 2l(r-1)}{\chi_{2l}^2/2l} $$
is known in the statistics literature as the $F$-random variable with $(2l(r-1),2l)$ degrees of freedom, whose CDF can be explicitly derived from its incomplete beta function representation (see, e.g., \cite[Chapter 26]{abramowitz+stegun}):
\ifconfver
\begin{align}
&\Pr( F_{2l(r-1),2l} \le \theta ) \nonumber \\
&= \frac{1}{((r-1)\theta+1)^{lr-1}} \sum_{j=l(r-1)}^{lr-1} {{lr-1} \choose j} (r-1)^j\theta^j, \,\,\, \theta \ge 0.\label{eq:F_cdf}
\end{align}
\else
\begin{align}
\Pr( F_{2l(r-1),2l} \le \theta ) & = \frac{1}{((r-1)\theta+1)^{lr-1}} \sum_{j=l(r-1)}^{lr-1} {{lr-1} \choose j} (r-1)^j\theta^j, \,\,\, \theta \ge 0.\label{eq:F_cdf}
\end{align}
\fi
The desired result then follows from \eqref{eq:gen_ellip_cdf} and \eqref{eq:F_cdf}.

\subsection{Proof of Proposition~\ref{prop:bing_rate}}
\label{sec:C}

By substituting the Bingham SBF equation \eqref{eq:w_bing} into the individual user rate \eqref{eq:C_SBF_i} and letting ${\bf A}_i = {\bf L} ( {\bf I}_N + P {\bf h}_i {\bf h}_i^H ) {\bf L}^H$,
the following rate expression is obtained:
\ifconfver
\begin{align}
& \mathbb{E}_{\bm{w}}[ \log( 1 + P | {\bf h}_i^H {\bm w} |^2 ) ] \nonumber\\
&= \mathbb{E}_{\bm{\alpha}} [ \log( \bm{\alpha}^H {\bf A}_i \bm{\alpha} ) ] - \mathbb{E}_{\bm{\alpha}} [ \log( \bm{\alpha}^H {\bf L} {\bf L}^H \bm{\alpha} ) ],\label{eq:C_eq1}
\end{align}
\else
\begin{align}
 \mathbb{E}_{\bm{w}}[ \log( 1 + P | {\bf h}_i^H {\bm w} |^2 ) ]
&= \mathbb{E}_{\bm{\alpha}} [ \log( \bm{\alpha}^H {\bf A}_i \bm{\alpha} ) ] - \mathbb{E}_{\bm{\alpha}} [ \log( \bm{\alpha}^H {\bf L} {\bf L}^H \bm{\alpha} ) ],\label{eq:C_eq1}
\end{align}
\fi
where, we recall, $\bm{\alpha} \sim \mathcal{CN}( {\bf 0}, {\bf I}_r )$.  Consider the spectral decompositions ${\bf A}_i = {\bf U} {\bf D} {\bf U}^H$ and ${\bf L}{\bf L}^H = {\bf Q} \bm{\Lambda} {\bf Q}^H$,
where ${\bf U}$ and ${\bf Q}$ are unitary, and ${\bf D}$ and $\bm{\Lambda}$ are diagonal whose diagonal elements are the eigenvalues of ${\bf A}_i$ and ${\bf L}{\bf L}^H$, resp.
Let $\mu_{i,1},\ldots,\mu_{i,r}$ be the diagonal elements of ${\bf D}$,
and $\lambda_1,\ldots,\lambda_r$ be the diagonal elements of $\bm{\Lambda}$.
By further letting $\bm{\alpha}^\prime = {\bf U}^H \bm{\alpha} \sim \mathcal{CN}( {\bf 0}, {\bf I}_r )$ and $\bm{\alpha}^{\prime\prime} = {\bf Q}^H \bm{\alpha} \sim \mathcal{CN}( {\bf 0}, {\bf I}_r )$,
we can rewrite \eqref{eq:C_eq1} as
\begin{align}
& \mathbb{E}_{\bm{w}}[ \log( 1 + P | {\bf h}_i^H {\bm w} |^2 ) ] \nonumber \\
& = \mathbb{E}_{\bm{\alpha}^\prime} \left[ \left( \sum_{k=1}^r \mu_{i,k} | \alpha^\prime_k |^2   \right) \right]
    - \mathbb{E}_{\bm{\alpha}^{\prime\prime}} \left[ \left( \sum_{k=1}^r \lambda_k | \alpha^{\prime\prime}_k |^2   \right) \right] \nonumber \\
&= \varphi( \bm{\mu}_i ) - \varphi( \bm{\lambda} ), \label{eq:C_eq2}
\end{align}
where $\varphi$ has been defined in \eqref{eq:varphi}.
One can then deduce from ${\bf W}^\star = {\bf L}^H {\bf L}$ that
$\lambda_1,\ldots,\lambda_r$ are also the positive eigenvalues of ${\bf W}^\star$.
Since we have, in addition, ${\rm Tr}({\bf W}^\star) = 1$
(as implied by the structure of (MC)), we get
\begin{align}
& \bm{\mu}^T_i {\bf 1} = {\rm Tr}( {\bf A}_i ) = {\rm Tr}({\bf L} ( {\bf I}_N + P {\bf h}_i {\bf h}_i^H ) {\bf L}^H) \nonumber \\
&= {\rm Tr}( {\bf W}^\star ( {\bf I}_N + P {\bf h}_i {\bf h}_i^H ) ) = 1 + P \rho_i.
\label{eq:C_eq3}
\end{align}
Upon substituting \eqref{eq:C_eq3} into \eqref{eq:C_eq2}, we obtain the result claimed in Proposition~\ref{prop:bing_rate}.

\subsection{Proof of Theorem \ref{thm:sdr_alam_approx}}
\label{sec:sdr_alam_approx_pf}

Consider a fixed $j\in\{1,\ldots,L\}$ in Algorithm \ref{alg1} and let $\tilde{\mathbf{W}} = \tilde{\mathbf{B}}_j\tilde{\mathbf{B}}_j^H$.  The proof consists of four steps:

\noindent\emph{Step $1$: }
For any $\bm{\mu} \in \mathbb{C}^N$, we have $\bm{\mu}^H\bm{\xi}_i \sim \mathcal{CN}(0,\bm{\mu}^H\mathbf{W}^\star\bm{\mu})$ and ${\mbox{Tr}}(\tilde{\mathbf{W}}\bm{\mu}\bm{\mu}^H)=\frac{1}{2}\sum_{i=1}^{2}|\bm{\mu}^H\bm{\xi}_i|^2$.  Hence, following \cite[Proposition A5.5]{inbook:So_bokcha}, for any $\beta \in (0, 1)$,
\begin{eqnarray}\label{pro0}
\Pr\Big( {\mbox{Tr}}(\tilde{\mathbf{W}}\bm{\mu}\bm{\mu}^H) \le \beta  \mbox{Tr} (\mathbf{W}^{\star}\bm{\mu}\bm{\mu}^H)\Big) \le e^{2(1-\beta +\ln \beta )}.
\end{eqnarray}

\noindent\emph{Step $2$: } Let $\mathbf{W}^\star = \mathbf{U}{\bm\Lambda}\mathbf{U}^H$ be the spectral decomposition of $\mathbf{W}^{\star}$. Observe that ${\mbox{Tr}}(\tilde{\mathbf{W}})=\frac{1}{2}\sum_{i=1}^{2}||\bm{\xi}_i||^2 \sim \frac{1}{2}\sum_{i=1}^{2}||\bm{\eta}_i||^2$, where $\bm{\eta }_i \sim \mathcal{CN}(\mathbf{0}, {\bm\Lambda})$ and $\bm{\eta }_1$, $\bm{\eta }_2$ are independent. Moreover, we have $\frac{1}{2}\sum_{i=1}^{2}||\bm{\eta}_i||^2 = \frac{1}{2}\sum_{j=1}^{N}\sum_{i=1}^{2}|\eta_{ij}|^2$, where $\eta_{ij} \sim \mathcal{CN}(0, \Lambda_{jj})$, and $\{\eta_{ij}\}$ are independent. Thus, for any $\alpha \in (1, \infty)$,
\begin{align*}
& \Pr\Big( {\mbox{Tr}}(\tilde{\mathbf{W}}) \ge \alpha \mbox{Tr} (\mathbf{W}^{\star})\Big) \\
&= \Pr\left( \frac{1}{2}\sum_{j=1}^{N}\sum_{i=1}^{2}|\eta_{ij}|^2 \ge \alpha \sum_{j=1}^{N} \Lambda_{jj}\right)\\
&= \Pr\left(\sum_{j=1}^{N} \Lambda_{jj} \sum_{i=1}^{4}|{\tilde\eta}_{ij}|^2 \ge \alpha \sum_{j=1}^{N} \Lambda_{jj}\right),
\end{align*}
where ${\tilde\eta}_{ij} \sim \mathcal{N}(0, 1/4)$. Now, using the argument in the proof of \cite[Proposition 2.1]{Jnl:SoYe2008} (see the remark after the proof of \cite[Proposition 2.2]{Jnl:SoYe2008}), we see that for $\alpha \ge 4/3$,
\begin{align}\label{pro2}
\Pr\Big( {\mbox{Tr}}(\tilde{\mathbf{W}}) \ge \alpha \mbox{Tr} (\mathbf{W}^{\star})\Big)\le e^{-\frac{1}{2}(\alpha+4\ln\frac{3}{4})}.
\end{align}

\noindent\emph{Step $3$: }
By setting $\beta = (e\sqrt{2.4M})^{-1}$ and $\alpha = 2\ln(2.4)-4\ln(3/4)\approx 2.902$ in \eqref{pro0} and \eqref{pro2}, resp., we obtain
\begin{align}
 \Pr\Big( {\mbox{Tr}}(\tilde{\mathbf{W}}\mathbf{h}_i\mathbf{h}_i^H) \le \beta {\mbox{Tr}} (\mathbf{W}^{\star}\mathbf{h}_i\mathbf{h}_i^H)\Big) &\le \frac{1}{2.4M} \quad\forall i, \label{eq:subopt} \\
\noalign{\smallskip}
 \Pr\Big( {\mbox{Tr}}(\tilde{\mathbf{W}}) \ge \alpha \mbox{Tr} (\mathbf{W}^{\star})\Big) &\le \frac{1}{2.4}. \label{eq:power}
\end{align}
Consider now the events
\begin{align*}
E &= \big\{ {\mbox{Tr}}(\tilde{\mathbf{W}}\mathbf{h}_i\mathbf{h}_i^H) \ge \beta {\mbox{Tr}} (\mathbf{W}^{\star}\mathbf{h}_i\mathbf{h}_i^H) \,\,\,\mbox{for }i=1,\ldots,M \big\}, \\
F &= \big\{ {\mbox{Tr}}(\tilde{\mathbf{W}}) \le \alpha \mbox{Tr} (\mathbf{W}^{\star}) \big\}.
\end{align*}
Using \eqref{eq:subopt}, \eqref{eq:power} and the union bound, we compute
$$ \Pr( E \cap F ) \ge \frac{1}{6}. $$
In particular, with probability at least $1/6$, we have
$$ \frac{{\mbox{Tr}}(\tilde{\mathbf{W}}\mathbf{h}_i\mathbf{h}_i^H)}{{\mbox{Tr}}(\tilde{\mathbf{W}})} \ge \frac{\beta}{\alpha} \cdot \frac{{\mbox{Tr}} (\mathbf{W}^{\star}\mathbf{h}_i\mathbf{h}_i^H)}{\mbox{Tr} (\mathbf{W}^{\star})} \ge \frac{{\mbox{Tr}} (\mathbf{W}^{\star}\mathbf{h}_i\mathbf{h}_i^H)}{12.22\sqrt{M}} $$
for $i=1,\ldots,M$ (recall that $\mbox{Tr}(\mathbf{W}^\star) = 1$).

\noindent\emph{Step $4$: } The result in Step 3 and the union bound imply that the event
$$ \left\{ \exists j: \frac{{\mbox{Tr}}(\mathbf{h}_i^H\tilde{\mathbf{B}}_j\tilde{\mathbf{B}}_j^H\mathbf{h}_i)}{{\mbox{Tr}}(\tilde{\mathbf{B}}_j\tilde{\mathbf{B}}_j^H)} \ge \frac{{\mbox{Tr}} (\mathbf{W}^{\star}\mathbf{h}_i\mathbf{h}_i^H)}{12.22\sqrt{M}} \,\,\mbox{for } i=1,\ldots,M \right\} $$
occurs with probability at least $1-(5/6)^L$.  This, together with the construction of $\hat{\mathbf{B}}$ in Algorithm \ref{alg1}, implies the desired result.

\bibliographystyle{IEEEbib}
\bibliography{IEEEabrv,ref}

\end{document}